\documentclass[sigconf]{acmart}

\AtBeginDocument{%
  }

\setcopyright{acmlicensed}
\copyrightyear{2026}
\acmYear{2026}
\setcopyright{cc}
\setcctype{by}
\acmConference[CHI '26]{Proceedings of the 2026 CHI Conference on Human Factors in Computing Systems}{April 13--17, 2026}{Barcelona, Spain}
\acmBooktitle{Proceedings of the 2026 CHI Conference on Human Factors in Computing Systems (CHI '26), April 13--17, 2026, Barcelona, Spain}
\acmPrice{}
\acmDOI{10.1145/3772318.3791259}
\acmISBN{979-8-4007-2278-3/2026/04}

\acmConference[CHI '26]{CHI Conference on Human Factors in Computing Systems}
\acmISBN{978-1-4503-XXXX-X/24/04}

\usepackage{tabu}
\usepackage{booktabs}
\usepackage{amsmath}
\usepackage{multirow}
\usepackage{gensymb}
\usepackage{tikz}
\usepackage{soul}
\usepackage{ulem}
\usepackage{tabularray}
\UseTblrLibrary{diagbox}
\usepackage{makecell}

\definecolor{figuregray}{RGB}{109,109,109}
\definecolor{myColor}{RGB}{47, 48, 48}

\newcommand{\alexin}[1]{{\textcolor{black}{#1}}}
\newcommand{\newalexin}[1]{{\textcolor{black}{#1}}}
\newcommand{\red}[1]{{\textcolor{black}{#1}}}
\newcommand{\revised}[1]{{\textcolor{black}{#1}}}

\newcommand*\component[1]{\tikz[baseline=(char.base)]{
            \node[shape=rectangle,fill=myColor,stroke=white, text=white, thick, inner sep= 1.5pt,minimum height=11pt,minimum width=11pt] (char) {\textsf{#1}}}}

\newcommand*\subcomponent[1]{\tikz[baseline=(char.base)]{
            \node[shape=rectangle,fill=myColor,text=white, thick, inner sep= 0.1pt,minimum height=12pt,minimum width=12pt] (char) {\textsf{\small #1}}}}

\newcommand{\name}[1]{\textit{BAIT}}

\begin{document}

\title{\name{}: Visual-illusion-inspired Privacy Preservation for Mobile Data Visualization}

\author{Sizhe Cheng}
\email{sizhe003@e.ntu.edu.sg}
\affiliation{%
  \institution{College of Computing and Data Science}
  \institution{Nanyang Technological University}
  \city{Singapore}
  \country{Singapore}
}
\authornote{Both authors contributed equally to this research.}

\author{Songheng Zhang}
\email{shzhang.2021@smu.edu.sg}
\affiliation{%
  \institution{School of Computing and Information Systems}
  \institution{Singapore Management University}
  \city{Singapore}
  \country{Singapore}
}
\authornotemark[1]

\author{Dong Ma}
\email{dm878@cam.ac.uk}
\affiliation{%
  \institution{Department of Computer Science and Technology}
  \institution{University of Cambridge}
  \city{Cambridge}
  \country{United Kingdom}
}

\author{Yong Wang}
\email{yong-wang@ntu.edu.sg}
\affiliation{%
  \institution{College of Computing and Data Science}
  \institution{Nanyang Technological University}
  \city{Singapore}
  \country{Singapore}
}
\authornote{Corresponding author.}

\begin{abstract}
\red{With the prevalence of mobile data visualizations, there have been growing concerns about their privacy risks, especially shoulder surfing attacks. Inspired by prior research on visual illusion, we propose BAIT, a novel approach to automatically generate privacy-preserving visualizations by stacking a decoy visualization over a given visualization.
It allows visualization owners at proximity to clearly discern the original visualization and makes shoulder surfers at a distance be misled by the decoy visualization, by adjusting different visual channels of a decoy visualization (e.g., shape, position, tilt, size, color and spatial frequency). We explicitly model human perception effect at different viewing distances to optimize the decoy visualization design. Privacy-preserving examples and two in-depth user studies demonstrate the effectiveness of BAIT in both controlled lab study and real-world scenarios.}
\end{abstract}




\maketitle

\section{Introduction}

With the wide usage of mobile devices like smartphones and tablets, it has become increasingly prevalent for people to access data visualizations on mobile devices. The visualizations displayed on such mobile devices are often called \textit{mobile data visualizations}~\cite{lee2021mobile}.
Given that mobile devices can be used anytime and anywhere, one of the most common threats to mobile data visualization is ``shoulder surfing attacks''~\cite{abdrabou2022understanding}, where the unauthorized third parties nearby, i.e., \textit{shoulder surfers}, can easily see the mobile data visualizations.
Prior studies have shown that such kinds of shoulder surfing attacks often happen in public areas like coffee shops, restaurants and subways~\cite{wiedenbeck2006design} and are performed without the awareness of victims~\cite{eiband2017understanding,ali2014protecting}.
What makes matters worse is that mobile data visualizations are often used to show sensitive personal information like personal health, finance and travel data~\cite{lee2018data}.
\red{Visualizations can easily expose sensitive information through a brief glance, such as financial trends and personal health data in mobile apps} ~\cite{Kerren2008Information}.
Therefore, 
the methods to achieve privacy-preserving mobile data visualizations are urgently demanded.

\red{However, compared with other forms of digital information like texts, the protection of data visualizations from unauthorized viewing presents unique challenges. Effective data visualizations leverage the human brain's innate capacity for rapid pattern recognition. Visual perception operates as a high-bandwidth parallel processor, enabling viewers to detect meaningful patterns in less than 500 milliseconds, far faster than text-based information processing~\cite{few2013information}. This efficiency stems from preattentive visual attributes like color, position, and size, which allow well-designed data visualizations to communicate critical insights at a glance~\cite{ware2013information, blascheck2021characterizing}. However, this design philosophy creates an inherent vulnerability: the same mechanisms that enable legitimate users to quickly extract insights also allow unauthorized observers to rapidly acquire sensitive information through a brief and casual glance.}

Existing methods that can be used to enhance privacy preservation of mobile data visualizations consist of two groups: hardware-based and software-based methods.
The hardware-based methods mainly refer to the various privacy films that can be attached to mobile device screens and prevent shoulder surfers from viewing mobile data visualizations.
But they come with intrinsic limitations~\cite{ali2014protecting}, such as negative effect on the color quality, decreased screen sensitivity, additional cost and the incapability of preserving privacy from many viewpoints.
\alexin{Software-based methods have been developed to address the limitations inherent to hardware-based approaches. Prior research has investigated the privacy preservation of grayscale images and text on mobile devices~\cite{papadopoulos2017illusionpin,chen2019keep}. Such approaches are tailored to the needs of different specific mobile applications. Other studies explored the privacy preservation of photos, which, however, is at the expense of compromising the necessary image details~\cite{von2016you}.}

\red{Fundamentally, these conventional approaches are misaligned with the nature of visualization information processing. Unlike text-based information that requires sequential reading and cognitive processing, a shoulder surfer can quickly discern trends from line charts, outliers in a bar chart, or proportional distributions from pie charts without sustained attention or deliberate examination~\cite{zhang2023don}.} \red{This characteristic of data visualization, what researchers term "glanceable visualization"~\cite{blascheck2021characterizing}, renders traditional privacy protection strategies that rely on slowing down information acquisition largely ineffective. Conventional approaches such as text obfuscation or gradual revelation become inadequate when core insights can be extracted from brief visual exposures lasting mere seconds.}
Most recently, Zhang \textit{et al}.~\cite{zhang2023don} 
transform colorful mobile data visualizations into privacy-preserving ones by adjusting their spatial frequency and luminance contrast. However, it
relies on manual configurations and is not easy to achieve an ideal trade-off between the visualization visibility at proximity and privacy preservation at a far distance.

In this paper,
we develop \name{}, a novel \textit{fully-automated} approach to achieve \underline{B}etter privacy preserv\underline{A}tion for mob\underline{I}le data visualiza\underline{T}ion, \red{which exploits visual illusion~\cite{gregory1997knowledge,andresyuk2024visual}} that \red{can mislead people's visual judgment}.
Specifically,
we propose to overlay a decoy visualization over the original mobile data visualization to camouflage it and distract shoulder surfers' attention from the original visualization to the decoy visualization.

\begin{figure}[t]
    \centering
    \includegraphics[width=0.8\columnwidth]{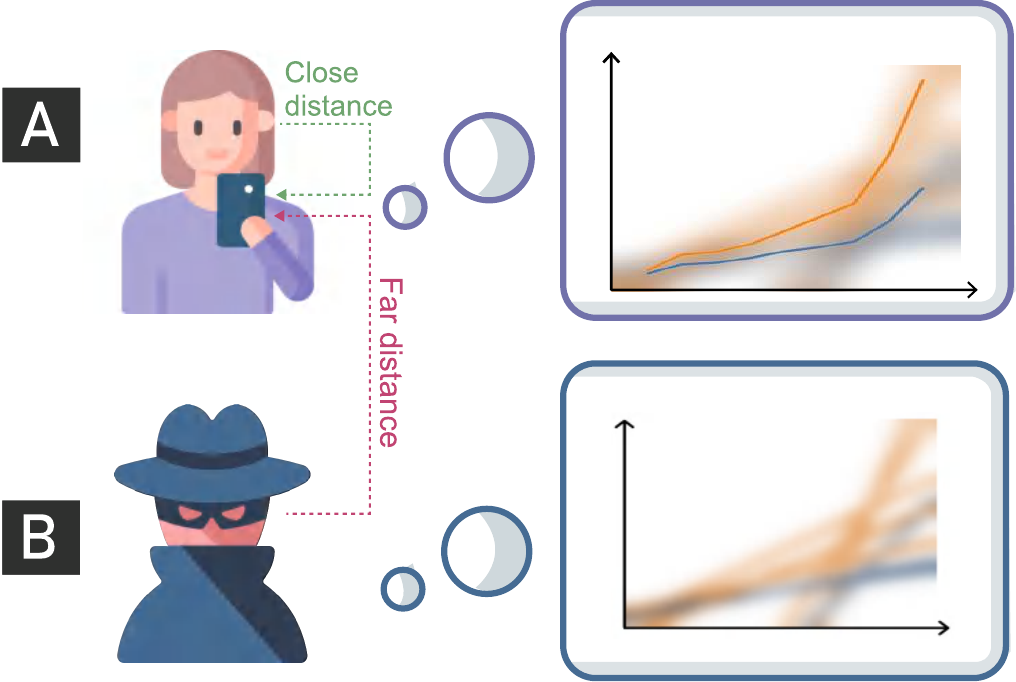}
    \caption{Illustration of our method. \protect\component{A} A user can discern the original visualization content from the privacy-preserving visualization at a close viewing distance. \protect\component{B}  At a far distance, a shoulder surfer perceives only the decoy visualization from the privacy-preserving visualization.}
    \label{fig: bait_illustration}
\end{figure}

\newalexin{Informed by prior research on psychology~\cite{rosli2015gestalt}, human vision system~\cite{connor2004visual} and visualization design~\cite{munzner2014visualization}, we systematically \red{summarize the visual variables} of decoy visualization, enabling the identification of essential visual channels, such as shape and position, which are critical for crafting effective decoys. Building upon these \red{visual variables}, \name{} introduces a framework for the creation of decoy visualizations. This framework is designed to subtly alter visual channels in the decoy visualization. It ensures that they are visually similar to, yet distinct in information content from, the original visualization, thereby misleading the shoulder surfers. Then, the privacy-preserving visualization is created by superimposing the decoy and original visualizations.}
\name{} explicitly models the perception similarity between the final privacy-preserving visualization and the original visualization (or decoy visualization), and evaluates the difference between the perception similarity at proximity and the perception similarity at a distance. 
\newalexin{By optimizing such perception-driven model}, \name{} can guarantee good visibility of the original visualization for visualization owner~(Figure~\ref{fig: bait_illustration}~\component{A}) and effectively prevent shoulder surfers from viewing the original visualization~(Figure~\ref{fig: bait_illustration}~\component{B}). 

We showcase privacy-preserving visualization examples of different charts to illustrate the results and usefulness of \name{}.
\red{Also, we conducted two carefully-designed user studies, involving 44 participants (32 in a controlled laboratory setting and 12 in a real-world field setting), to evaluate the privacy preservation effectiveness of \name{} and compare it with alternative methods as baselines.}
The result demonstrates the effectiveness of \name{} in generating privacy-preserving data visualizations and ensuring the visibility of original visualization for visualization owners.
The major contributions of this paper can be summarized as follows:
  
  
\begin{itemize}
  \item Based on interviews with 14 frequent visualization users, we identify and formalize three critical design implications for illusion-based privacy. These implications establish guidance for mobile privacy, enabling privacy protection to prioritize user experience while balancing security, filling the research gap that previously focused primarily on technical feasibility.
  
  \item We propose \name{}, a fully-automated computational approach that generates privacy-preserving visualizations. \name{} formulates the overlay of decoy visualizations as a constraint optimization problem based on Human Visual System (HVS) characteristics, ensuring a mathematically modeled balance between privacy and utility.
  
  \item We provide empirical evidence from two user studies ($N=44$) demonstrating that \name{} significantly outperforms baseline methods in mitigating shoulder surfing. Crucially, our results reveal that decoy-based visual masking can decouple the visualization owner's perception from the attacker's, challenging the conventional trade-off where privacy comes at the cost of data legibility. This informs future privacy-preserving interface design by validating that visualizations can be secured through perceptual manipulation rather than simple information degradation.
\end{itemize}

\section{Background}\label{sec: bg}

Before diving into the design of \name{}, we first provide 
\red{background knowledge} about \red{Visual Illusion} (Section~\ref{bg: decoy_effect}) and Human Vision System (Section~\ref{bg: hvs}).

\begin{figure}[ht!]
    \centering
    \includegraphics[width=0.8\linewidth]{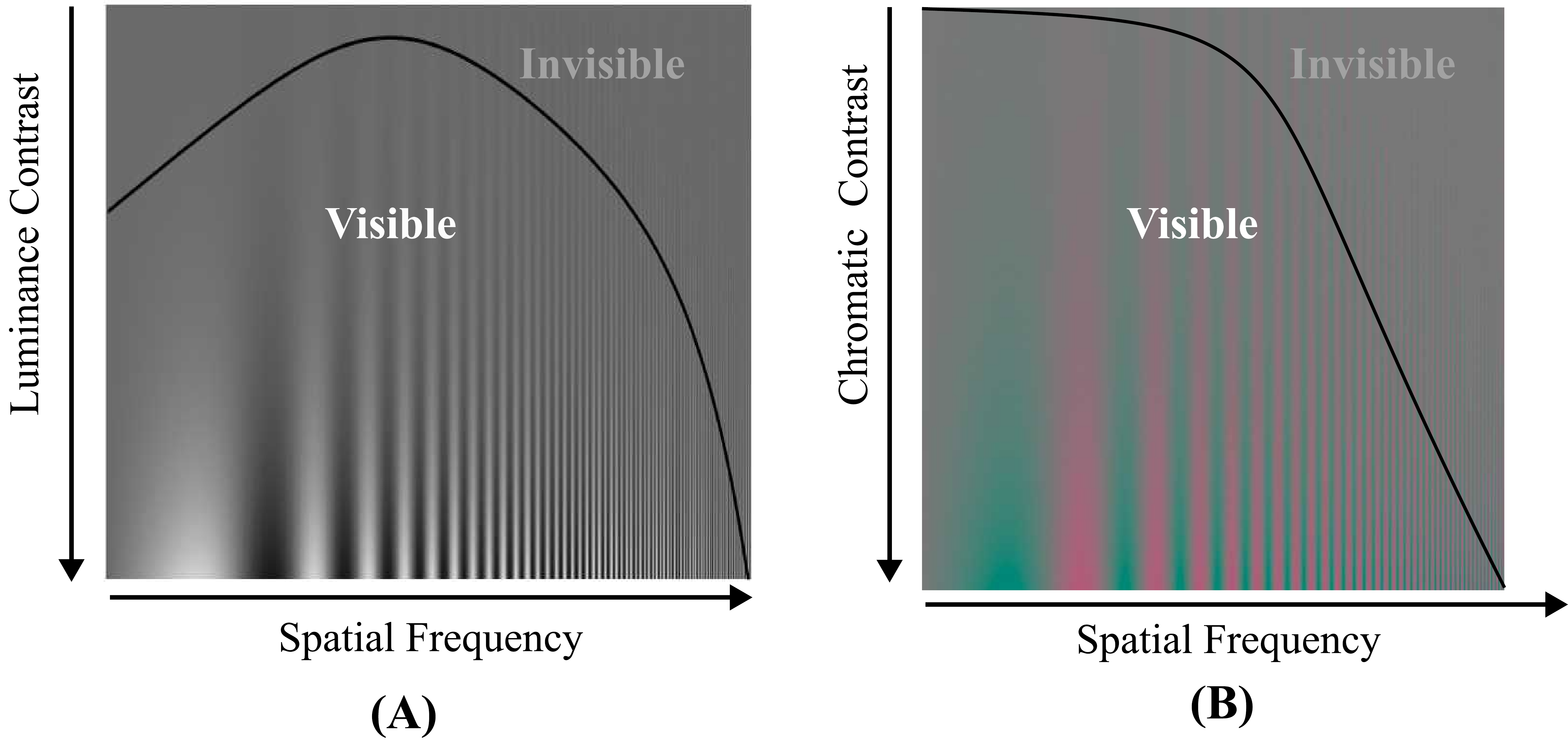}
    \caption{The contrast sensitivity function curve illustrates the relationship between spatial frequency, stimulus contrast, and human visual perception. (A) highlights the association between luminance contrast and spatial frequency, whereas (B) depicts the correlation between chromatic contrast and spatial frequency.}
    \label{fig: csf}
\end{figure}

\subsection{\red{Visual Illusion}}\label{bg: decoy_effect}

\red{Visual illusions are systematic deviations between visual perception and physical reality, where the brain's interpretation of visual information diverges from objective facts~\cite{gregory1997knowledge}. These phenomena result in erroneous perceptions, such as seeing curvature in straight lines or misjudging size and color, as famously demonstrated in the Müller-Lyer illusion where fin orientations cause identical line segments to be misperceived in length~\cite{andresyuk2024visual}. This occurs because visual perception operates as an active process of "hypothetical prediction" based on past experiences, rather than a passive recording of the world~\cite{gregory1997knowledge}. To quickly and efficiently understand the three-dimensional world we inhabit, the brain has evolved a set of interpretive shortcuts that rely on environmental cues such as context, lighting, and perspective. Visual illusions emerge precisely when these normally reliable shortcuts are deceived by specific patterns, revealing the deeper operating principles of the visual system~\cite{Franz2000Grasping}.}

\red{Consequently, visual illusions are not flaws in the visual system, but rather inherent byproducts of its efficient, predictive mechanisms that prioritize speed and utility over absolute accuracy. A crucial aspect of these illusions is their cognitive impenetrability; even when observers are consciously aware of an illusion's existence and understand the objective truth, the perceptual bias cannot be subjectively eliminated~\cite{kruglanski2018intuitive}. Our approach leverages this robust and predictable nature of visual illusion to create privacy-preserving visualizations by intentionally and systematically manipulating the perceptual judgments of shoulder surfers.}


\subsection{Human Vision System}\label{bg: hvs}
The Human Vision System (HVS) is crucial for capturing and processing information from the environment. A thorough understanding of the HVS is essential to determine whether humans can distinguish specific objects from others. The information processing in the HVS can be categorized into two parts: bottom-up and top-down processing~\cite{connor2004visual}.
Bottom-up processing is an unintentional mechanism influenced by an external stimulus in the environment, while top-down processing is a voluntary mechanism influenced by individuals' prior knowledge and contextual information to process visual information~\cite{connor2004visual}. Luminance contrast, chromatic contrast, and spatial frequency are key factors of the bottom-up processing in the HVS~\cite{panetta2008human,nothdurft2000salience, white2017colour, terzic2017texture}. Luminance contrast, as illustrated in Figure~\ref{fig: csf} (A), refers to the degree of brightness difference in an area (e.g., a pair of dark and bright bars). Chromatic contrast, depicted in Figure~\ref{fig: csf} (B), represents the colorfulness difference in an area (e.g., a pair of green and red bars). Spatial frequency, another significant aspect of the HVS, signifies the amount of detail presented in an image. For instance, as shown in Figure~\ref{fig: csf} (A), an image with more detailed areas (e.g., thinner pairs of dark and bright bars) has higher spatial frequencies.
The spatial frequency has coupling effects with luminance and chromatic contrasts~\cite{devalois1990spatial}, which determine the HVS's ability to identify different objects in an image. As the spatial frequency increases, the HVS requires higher contrasts to distinguish objects (i.e, a pair of bars), as demonstrated in Figure~\ref{fig: csf} (A) and (B). Furthermore, the HVS's perceived spatial frequency is not only determined by the intrinsic details in an image but also depends on the viewing distance~\cite{rovamo1992contrast}. As a human observes a fixed-size image from various distances, ranging from close to far, the spatial frequency of the visual information perceived by the HVS increases. This implies that for the same detailed area in an image, a person at a far viewing distance requires greater contrasts within that area than those at a close distance to identify the area clearly.
In contrast to bottom-up processing, top-down processing is related to the Gestalt Principle~\cite{gilbert2007brain}, which emphasizes that HVS perception of an object is not an assembly of its individual parts but rather a perception of the object as a whole. For example, people may have difficulty identifying two objects with similar appearances (e.g., shape, size, and orientation) as separate entities because the HVS tends to regard them as a single object. This phenomenon can be explained by the Gestalt Principle of Similarity~\cite{gilbert2007brain}, which leads the HVS to perceive two objects as a unified whole due to their shared visual characteristics. 
\red{\section{Formative Study: Understanding Privacy Concerns and Practices for Mobile Visualizations}}
\label{sec:formative_study}

\red{To ground our design in empirically-derived user needs rather than preliminary assumptions, we conducted a formative interview study with 14 participants (P1--P14). Below we report our study goals, design, and key findings based on participants' real feedback, and key findings that directly informed \name{}.}

\red{\subsection{Research Goals and Questions}}
\red{The study aimed to understand where and how people view visualizations on mobile devices, what they perceive as privacy risks, how existing countermeasures perform, and what they want from a privacy-preserving solution. We organized the inquiry around the following research questions:}

\begin{itemize}
\item[\textbf{RQ1:}] \red{How prevalent and significant is the shoulder surfing threat when people view data visualizations on mobile devices?}
\item[\textbf{RQ2:}] \red{What are the pros and cons of existing anti-peeping methods, and are they suitable for protecting mobile visualizations?}
\item[\textbf{RQ3:}] \red{What are the key user requirements for a tool designed to protect mobile visualizations from shoulder surfing?}
\end{itemize}

\red{\subsection{Study Design}}
\red{Our participants were recruited from a local university and included both current students and recent graduates who frequently use visualizations on mobile devices (inclusion criterion: more than two times per week). \revised{The average age was 24 (min 21, max 30, SD 3), and participants included 7 men and 7 women}. The research topics included: (i) contexts and frequency of viewing mobile visualizations, (ii) perceived privacy risks and concrete scenarios, (iii) current coping behaviors and tools (e.g., privacy screen protectors), (iv) chart types and elements perceived as sensitive or easily gleaned, and (v) the ideal ``privacy mode'' characteristics. Participants discussed real experiences across public and semi-public settings (e.g., open offices, subways, cafés, conferences, client meetings, lab visits). Each interview lasted approximately 30 minutes, and participants who completed the session received a compensation of \$10 for their time. With consent, we collected and analyzed responses using thematic analysis, and we report frequency counts to indicate the prevalence when appropriate.}

\red{\subsection{Key Findings}}

\red{We organize our findings to directly correspond with the research questions outlined above.}

\textbf{For RQ1: Prevalence and Significance of the Threat.}
\red{Our findings indicate that the threat of shoulder surfing is both prevalent and significant enough to alter user behavior. A majority of participants (11/14) explicitly expressed concerns about their on-screen visualizations being viewed in public or semi-public settings (P1-3, P6, P8-14). The threat is tangible and salient for professional or in-progress content; participants reported reducing viewing frequency or deferring sensitive charts in open offices, client meetings, conferences, and transit, with heightened worry when potential viewers are close colleagues due to professional risks (supported by P2, P3, P6, P8, P9, P11-14).}
\begin{quote}
\red{``When you're making evaluation charts, whether the performance is good or bad, it's related to your research. In a public place... you're still a bit concerned about it being seen... the frequency [of viewing] would indeed decrease.''} \red{(P8)}
\end{quote}
\begin{quote}
\red{``In an open office, being seen by a colleague, the concern is about company internal information; on the subway, being seen by a stranger, the concern is about personal privacy and security. The former's pressure is more of a professional risk.''} \red{(P11)}
\end{quote}
\red{The significance of the threat is also tied to content, with financial charts (e.g., stock performance) and professional data (e.g., project progress) being most frequently cited as sensitive.}

\textbf{For RQ2: Suitability of Existing Countermeasures.}
\red{Our investigation into existing methods revealed that they are largely unsuitable for protecting data visualizations. Adoption of privacy screen protectors was low, with only 3 participants having used them previously (P6, P10, P14) and 3 participants currently using them (P9, P11, P12). Of the six participants with experience, five reported significant usability drawbacks, including reduced brightness, color distortion, and eye strain (P6, P9-11, P14), which critically degrade the readability of visualizations.}
\red{Furthermore, these countermeasures fail to address the core challenge: the inherent glanceability of visualizations. Many participants emphasized that proportions in pie charts and trends in line charts can be understood in an instant, a problem current tools do not solve.}
\begin{quote}
\red{``For large-scale scientific computing, data files are unlikely to be directly spied upon... Visualization content is another matter; it contains a large amount of useful information, and is also very easy to glimpse and remember, at least much simpler than remembering 200 numbers.''} \red{(P1)}
\end{quote}
\begin{quote}
\red{``Sometimes I also have the need to look at my phone from different angles. You can't use a privacy film 24/7. Otherwise, when I want to look at my phone, I always need to adjust it to a direct angle.''} \red{(P14)}
\end{quote}

\textbf{For RQ3: User Requirements for a Privacy-Preserving Tool.}
\red{When asked what they want from a solution, users described a clear need for a tool that is \textbf{effortless, readable, and customizable}. Priorities include one-touch/automatic activation and flexible control over what to protect (P1, P6, P8-P11). The foremost requirement is that the user's own ability to read the chart must not be compromised.}
\begin{quote}
\red{``I must be able to see and understand it clearly, otherwise, it would be putting the cart before the horse. Then, ease of operation is also important; it must be a one-click solution. If it's too complicated to operate, I won't use it.''} \red{(P13)}
\end{quote}
\begin{quote}
\red{``It should automatically detect the current environment (like a public place) and then blur or hide sensitive values... show a simplified version to the outside... while I can still see the full information through detailed interaction.''} \red{(P11)}
\end{quote}
\red{When asked which chart types are most easily compromised by a glance, participants most often named \textbf{pie charts} (large, distinct color blocks; P2, P3, P6, P8, P9, P11, P14) and \textbf{line charts} (immediately understandable trends; P1, P3, P5, P8, P11, P13, P14). This suggests a key requirement is to specifically target these easily glanceable visual channels.}

\red{\subsection{Design Implications for \name{}}}
\red{Based on the findings above, we distilled three core design implications that guided the development of \name{}, each responding to insights from our research questions.}
\begin{itemize}
    \item \red{\textbf{DI1 -- Context-Aware and Low-Friction control.} This implication addresses findings from \textbf{RQ1} and \textbf{RQ3}. The fact that users alter their behavior in specific high-risk contexts (RQ1) highlights the need for a solution that can be easily engaged in those moments. This is reinforced by the explicit user requirement for simple, "one-click" controls (RQ3) to reduce the friction of seeking protection.}
    \item \red{\textbf{DI2 -- Disrupt glanceability.} This implication is a direct response to findings from \textbf{RQ2} and \textbf{RQ3}. Our analysis showed that existing countermeasures fail because they do not mitigate the inherent glanceability of visualizations (RQ2). Concurrently, users identified that the most easily gleaned information comes from proportions (pie charts) and trends (line charts) (RQ3). Therefore, a successful solution must target these specific visual channels.}
    \item \red{\textbf{DI3 -- Ensure high usability without sacrificing the user's task.} This implication stems directly from findings in \textbf{RQ2} and \textbf{RQ3}. The widespread criticism of hardware films for degrading brightness and color (RQ2) makes it clear that usability is paramount. This aligns with the primary user requirement that any privacy solution must preserve the owner's ability to clearly see and understand their own data (RQ3).}
\end{itemize}

\begin{figure*}[ht]
    \centering
    \includegraphics[width=0.95\linewidth]{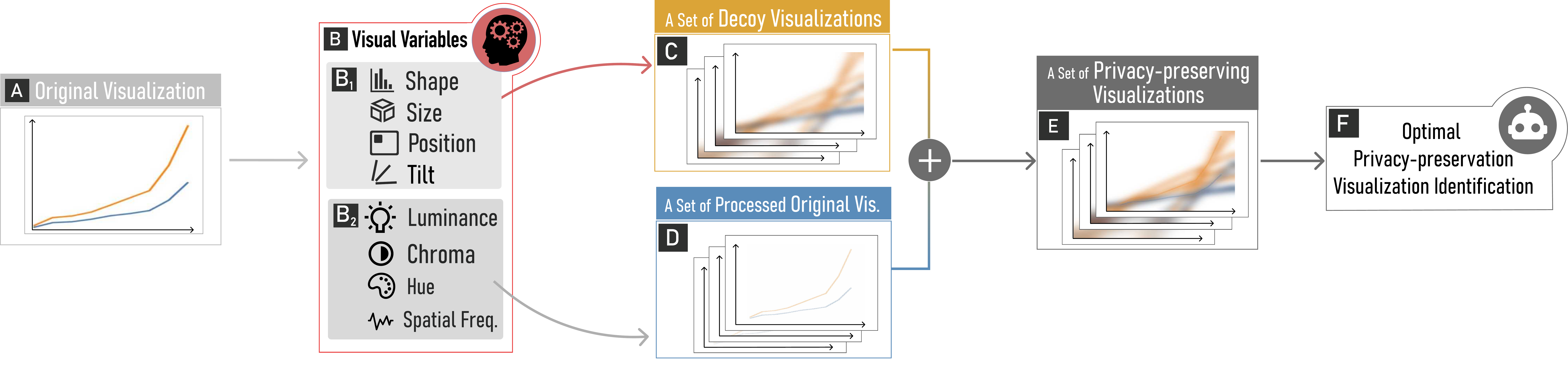}
    \caption{
    Method Overview: \protect\component{A} Input: Original visualization. \protect\component{B} \red{Visual variables} in our privacy-preserving visualization. \protect\subcomponent{B$_1$} 
    Visualization-dependent channels.
    \protect\subcomponent{B$_2$} 
    Visualization-agnostic channels.
    \protect\component{C} A set of decoy visualizations generated based on the \red{visual variables} in both \protect\subcomponent{B$_1$} and \protect\subcomponent{B$_2$}. \protect\component{D} A set of processed original visualizations generated based on the varying values of a channel: spatial frequency in \protect\subcomponent{B$_2$}. \protect\component{E} A collection of privacy-preserving visualizations created by stacking the set of decoy visualizations and the set of processed original visualizations together. \protect\component{F} A perception-driven model that efficiently identifies and selects the most effective privacy-preserving visualization from the collection, ensuring optimal balance between visibility and privacy.
     }
    \label{fig: method_overview}
\end{figure*}
\section{Method}
In this section, we introduce \name{}, an innovative method for enhancing privacy in mobile data visualizations through superimposing a strategically crafted decoy visualization over an original visualization.
Figure~\ref{fig: method_overview} shows an overview of our approach.
Our approach first summarized a set of visual variables for the decoy, including shape, position, and color (Section \ref{method_design_space}). We then propose a framework to generate these decoys (Section \ref{method_decoy_vis_generation}) and a perception-driven model that uses image similarity metrics to find the optimal balance between owner visibility and attacker privacy (Section \ref{method_optimization}).

\begin{figure}[ht!]
    \centering
    \includegraphics[width=\linewidth]{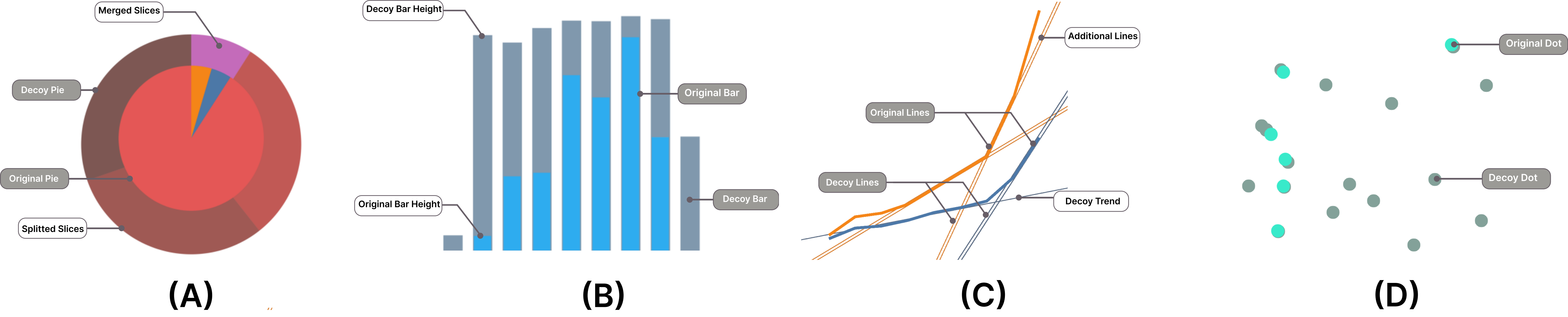}
    \caption{Illustrations of decoy visualization generation for four popular visualization types. \protect\component{A} Bar chart: Manipulating the number and height of bars to obscure the original visualization; \protect\component{B} Line chart: Modifying the decoy lines' position, length, and tilt; \protect\component{C} Scatter plot: Changing the decoy circles' position and number; \protect\component{D} Pie chart: Altering the angles of decoy slices and enlarging the decoy pie chart.}
    \label{fig: method_decoy_data}
\end{figure}

\subsection{\red{Visual Variables} of \name{}}
\label{method_design_space}

The decoy visualization camouflages the original visualization from shoulder surfers. Creating an effective decoy presents three challenges: it must be similar enough to the original to confuse an observer yet distinct enough for the owner; it must be adaptable to common visualization types; and its generation must be automatic, \red{which directly responds to the need for context-aware and low-friction control identified in DI1}. To address these challenges, we define a set of \red{visual variables} for decoy design, drawing from prior work in perception~\cite{connor2004visual,huber1982adding, rosli2015gestalt} and visualization~\cite{munzner2014visualization, senay1994knowledge}.
Specifically, visualizations consist of different visual marks and \textit{visual channels} controlling the appearance of visual marks~\cite{munzner2014visualization}.
Among all the visual channels, we identify five pivotal visual channels~\cite{munzner2014visualization}: shape, position, color, tilt, and size.
We also include spatial frequency, which affects visibility at different distances~\cite{isenberg2013hybrid, oliva2006hybrid}. 
By examining the dependency of visual channels on visualization types, we categorize all the visual channels in our \red{visual variables} into two groups: \textit{visualization-dependent channels}~(Figure~\ref{fig: method_overview}~\component{B$_1$}) and \textit{visualization-agnostic channels}~(Figure~\ref{fig: method_overview}~\component{B$_2$}).

Visualization-dependent channels include \textbf{Shape}, \textbf{Position}, \textbf{Tilt}, and \textbf{Size}. The utility of these channels varies by visualization type, and understanding them is paramount for decoy design. \textbf{Shape} defines the visual marks themselves. \textbf{Position} is a dominant channel for encoding quantitative data~\cite{heer2010crowdsourcing,cleveland1984graphical}. \textbf{Tilt}, or the angle of a mark, is often used to represent trends in line charts~\cite{munzner2014visualization}. \textbf{Size} is another expressive channel for quantitative variables, referring to length, area, or volume~\cite{iliinsky2011designing}. Manipulating these key channels is critical to creating an effective decoy that balances privacy and clarity.

The visualization-agnostic channels consist of Color and Spatial Frequency. \textbf{Color} properties—such as hue, luminance, and chroma—are key, as luminance and chromatic contrast substantially affect the HVS's ability to distinguish the original and decoy visualizations~\cite{borland2007rainbow, dastani2002role, bartram2017affective}. \textbf{Spatial Frequency} is also critical due to the HVS's distance-dependent perception: users at proximity perceive high-frequency details, while shoulder surfers at a farther distance~\cite{papadopoulos2017illusionpin} are more sensitive to low-frequency stimuli~\cite{oliva2006hybrid}. We leverage these channels to enhance privacy preservation.

\subsection{Decoy Visualization Generation}
\label{method_decoy_vis_generation}

Built upon the \red{visual variables} introduced in Section~\ref{method_design_space}, we propose a framework to generate detailed visualization choices to ensure privacy preservation.
Specifically, we first determine visualization-dependent channels such as shape, position, tilt
and size, and further adaptively adjust the visualization-agnostic channels of a decoy visualization.

\subsubsection{Rule-based Determination of Visualization-dependent Channels}
\label{method: initial_design_space}

We systematically surveyed prior studies on visual perception~\cite{panetta2008human, white2017colour, terzic2017texture} and privacy preservation~\cite{papadopoulos2017illusionpin, chen2019keep} and further determined the rules for designing 
the visualization-dependent channels of decoy visualizations.

\textbf{Use the same shape as that of the original visualization.}
In this study, our decoy visualizations primarily focus on the bar chart, line chart, scatter plot, and pie chart because they are the four most commonly used types of visualizations~\cite{battle2018beagle}.
Previous research has demonstrated the significant influence of Gestalt principles on human perception of visualizations~\cite{nesbitt2002applying}. In particular, the Law of Similarity indicates that when two visualizations share a similar visual appearance, humans tend to perceive them as a whole one~\cite{tran2021approaching}, thereby making it difficult to distinguish them.
Thus, the shape of visual marks in the decoy visualization should be the same as that of the original visualization, which can maximize their similarity and effectively confuse shoulder surfers.
For instance, for a bar chart, the decoy visualization is also a bar chart with different bars, and the decoy visualization for a line chart is another line chart as well.

\textbf{Randomize the position, tilt, and size channels of decoy visualization with adaptive constraints for different types of visualizations.}
After determining the shape channel of a decoy visualization, it is crucial to further determine its other visualization-dependent channels of the decoy visualizations, i.e., position, tilt, and size. Built upon prior visual perception research~\cite{todorovic2008gestalt, Herrmann2010WhenSM, Stuart2003TheOI}, our design goal is to generate a decoy that provides plausible yet misleading information to shoulder surfers while minimizing the perceptual impact on the visualization's owner. To this end, we first leverage image processing techniques~\cite{yuen1990comparative, leavers1993hough, suzuki1985topological} to extract these channels from the original visualization. We then randomize them within adaptive constraints designed for each specific type of visualization.

\revised{Specifically, for a \textbf{Bar chart}, we randomize the heights of the decoy bars and ensure they are larger than the original bars (Figure~\ref{fig: method_decoy_data} \component{A}) to distract shoulder surfers from the original bar chart, while maintaining the original bar chart's visibility to owners. For a \textbf{Line chart} (Figure~\ref{fig: method_decoy_data} \component{B}), we use the Hough Transform~\cite{duda1972use} to detect the original line chart's tilts and positions, and make the decoy lines' tilts and positions partially aligned with the original lines, while slightly adjusting the decoy lines' length, resulting in a visually similar but misleading trend in the decoy line chart. For a \textbf{Scatter plot} (Figure~\ref{fig: method_decoy_data} \component{C}), we extract original dot coordinates using a Circle Hough Transform~\cite{Pedersen2009CircularHT} and then generate new decoy dots by displacing each within a pre-set deviation range, making them close but not identical to the original. Finally, for a \textbf{Pie chart} (Figure~\ref{fig: method_decoy_data} \component{D}), we split large original slices into smaller ones and merging small ones into larger ones; we also make the decoy's radius larger than the original, leveraging the normalization model of attention~\cite{Herrmann2010WhenSM} to increase its visual dominance and distract shoulder surfers.}
\red{This strategy of altering data-encoding channels like size, position, and tilt directly addresses DI2, as it is designed to disrupt shoulder surfers' accurate perception of the original data visualization, especially on the crucial insights of original data visualizations like trends, proportions and clustering effect that users identified as being quite vulnerable to even a quick glance~\cite{zhang2023don}.}

\subsubsection{HVS-based Adjustment of Visualization-agnostic Channels}
\label{method: fine_tune}
\label{method-visual-property-adjustment}

As discussed in Section~\ref{method_design_space}, visualization-agnostic channels like color and spatial frequency are universal across different types of visualizations and can affect human perception of visual stimulus. The color channel can be further categorized into color hue, luminance, and chroma.
Thus, we also adjust the color and spatial frequency to achieve optimal privacy preservation. Such an adjustment is guided by the characteristics of HVS (Section~\ref{bg: hvs}).

To adjust the color of the decoy visualization, we first decompose the original visualization's color into the LCH space~\cite{berns1993mathematical},
 where the L, C, H channels correspond to the luminance, chroma, and hue of colors.
The detailed adjustment of the color of the decoy visualization is as follows:

\textbf{Set the color hue of the decoy visualization close to the original visualization.} 
According to the Gestalt principle of Similarity~\cite{gilbert2007brain}, it is more difficult for the HVS to differentiate objects with similar colors. 
Therefore, to ensure the effectiveness of privacy preservation,
it is crucial to maintain hue similarity between the visual elements of decoy visualization and those of the original visualization.
For the original visualizations with
only one color hue,
we adopt the same color hue for the decoy visualization. 
When the original visualization has multiple types of color hues,
we set the color hue of the visual elements in the decoy visualization as the average hue of adjacent visual elements in the original visualization, ensuring the similarity of color hues with the original visualization. Such an average is calculated by using the circular mean method~\cite{jammalamadaka2001topics}, which is consistent with the human perception of colors. 

\textbf{Reduce the luminance contrast between decoy visualization and original visualization to an appropriate value.} 
In addition to color hue, the Human Vision System's capacity to discern objects is influenced by the luminance contrast (Section~\ref{bg: hvs}). Thus, we choose to modify the luminance values of the visual elements in the decoy visualization by manipulating the L (luminance) channel in the LCH color space. 

\textbf{Optimize the chromatic contrast between decoy visualization and original visualization.}
The chromatic contrast is another important factor that significantly influences the ability of the HVS in differentiating different objects~\cite{white2017colour}.
Thus, we adjust the chromatic contrast between the decoy visualization and the original visualization.
Similar to luminance contrast,
we only change the Chroma (C channel) of the decoy visualization and keep the original visualization's color value in the C channel unchanged.

For the \textbf{spatial frequency},
to exploit the HVS's distance-dependent sensitivity to spatial frequencies (Section~\ref{bg: hvs}),
we \textbf{decrease the spatial frequency of decoy visualization} and \textbf{increase the spatial frequency of the original visualization}.
In particular, we employ a Gaussian filter to eliminate the high-frequency content~\cite{HUSSEIN201197} of the decoy visualization, which blurs the decoy visualization but preserves its overall visual appearance. 
Also, we follow the recent study~\cite{zhang2023don} and use a masking scheme to increase the spatial frequency of the original visualization, which processes both the major visual elements of a visualization (e.g., bars in a bar chart and lines in a line chart) and the corresponding auxiliary visual elements like x/y-axis and text labels.

\subsubsection{Perception-driven Combinatorial Optimization}\label{method_optimization}
\revised{\name{} employs a perception-driven optimization model to find the optimal combination of values for the visualization-agnostic channels. The goal is to strike an optimal balance between the visibility of the original visualization at a close distance and its privacy preservation at a far distance. This is achieved in two steps: distance-dependent rescaling and image similarity-based optimization.}

\revised{First, to model the influence of viewing distance, we simulate how the visualization is perceived by visualization owners at proximity and shoulder surfers at a distance. Since visual acuity decreases with viewing distance and perceived object size shrinks as the visual angle narrows~\cite{levin1993visual}, we use a downsampling technique~\cite{gu2013self} to generate two versions of any visualization $I$: a close-distance version $I_{c}$ and a far-distance version $I_{f}$. This process is governed by a downsampling factor $\gamma$, which is calculated based on viewing distance $D$ and human visual angles~\cite{gu2015quality}.}

\revised{Second, we find the optimal privacy-preserving visualization $I^{p}$ by maximizing its similarity to the original ($I^{o}$) and decoy ($I^{d}$) visualizations at these two distances. The optimal $I^{p}$ should satisfy two conditions: (1) at close distance, its perceived version $I^{p}_{c}$ should have high similarity to the original $I^{o}_{c}$; (2) at far distance, its perceived version $I^{p}_{f}$ should have high similarity to the decoy $I^{d}_{f}$.}

\revised{\textbf{Image similarity metrics.} We employ the Visual Saliency-based Index (VSI)~\cite{zhang2014vsi} to measure the similarity between $I^{p}$ and $I^{o}$, as VSI emphasizes regions that attract viewer attention (e.g., bars, lines, and data points in a visualization)~\cite{matzen2017data}. For the similarity between $I^{p}$ and $I^{d}$, we use the Multi-Scale Structural Similarity Index (MS-SSIM)~\cite{wang2003multiscale} to measure the similarity between the decoy visualization and privacy-preserving visualization~\cite{Panavas2023InvestigatingTV, abdrabou2022understanding}.}

\revised{Let $\mathrm{VSI}\left(A, B\right)$ denote the Visual Saliency-based similarity between images $A$ and $B$, and $\mathrm{SSIM}\left(A, B\right)$ denote the Multi-Scale Structural Similarity between images $A$ and $B$. The privacy-preserving visualization at close distance should be more similar to the original than at far distance, which is captured by:
\begin{equation}\label{formula: original_gap}
\mathrm{Gap}_{1} = \mathrm{VSI}\left(I^{o}_{c}, I^{p}_{c}\right) - \mathrm{VSI}\left(I^{o}_{f}, I^{p}_{f}\right).
\end{equation}
Conversely, the privacy-preserving visualization at far distance should be more similar to the decoy than at close distance:
\begin{equation}\label{formula: decoy_gap}
\mathrm{Gap}_{2} = \mathrm{SSIM}\left(I^{d}_{f}, I^{p}_{f}\right) - \mathrm{SSIM}\left(I^{d}_{c}, I^{p}_{c}\right).
\end{equation}
A higher $\mathrm{Gap}_{1}$ means the owner can clearly see the original content at close range, while a higher $\mathrm{Gap}_{2}$ means the shoulder surfer is more likely to perceive only the decoy at far range. We search for the optimal values of our four visualization-agnostic channels by maximizing:
\begin{equation}\label{formula: loss}
\alpha \cdot \mathrm{Gap}_{1} + \beta \cdot \mathrm{Gap}_{2},
\end{equation}
where $\alpha=\beta=0.5$. Detailed derivations are in Appendix B.}

\revised{We employ \textit{exhaustive search} to identify the optimal combination of our four visualization-agnostic channels: \textit{luminance}, \textit{chroma}, kernel size, and mask area~\cite{zhang2023don}. The latter two parameters control the spatial frequency of the decoy and original visualizations, respectively. This process generates a collection of candidates (Figure~\ref{fig: method_overview}~\component{E}), and we select the one with the highest value of Equation~\ref{formula: loss} (Figure~\ref{fig: method_overview}~\component{F}). The parameter ranges and search details are provided in Appendix B.}

\alexin{\section{Privacy-preserving Visualization Examples}}
\begin{figure*}[ht!]
    \centering
    \includegraphics[width=0.8\linewidth]{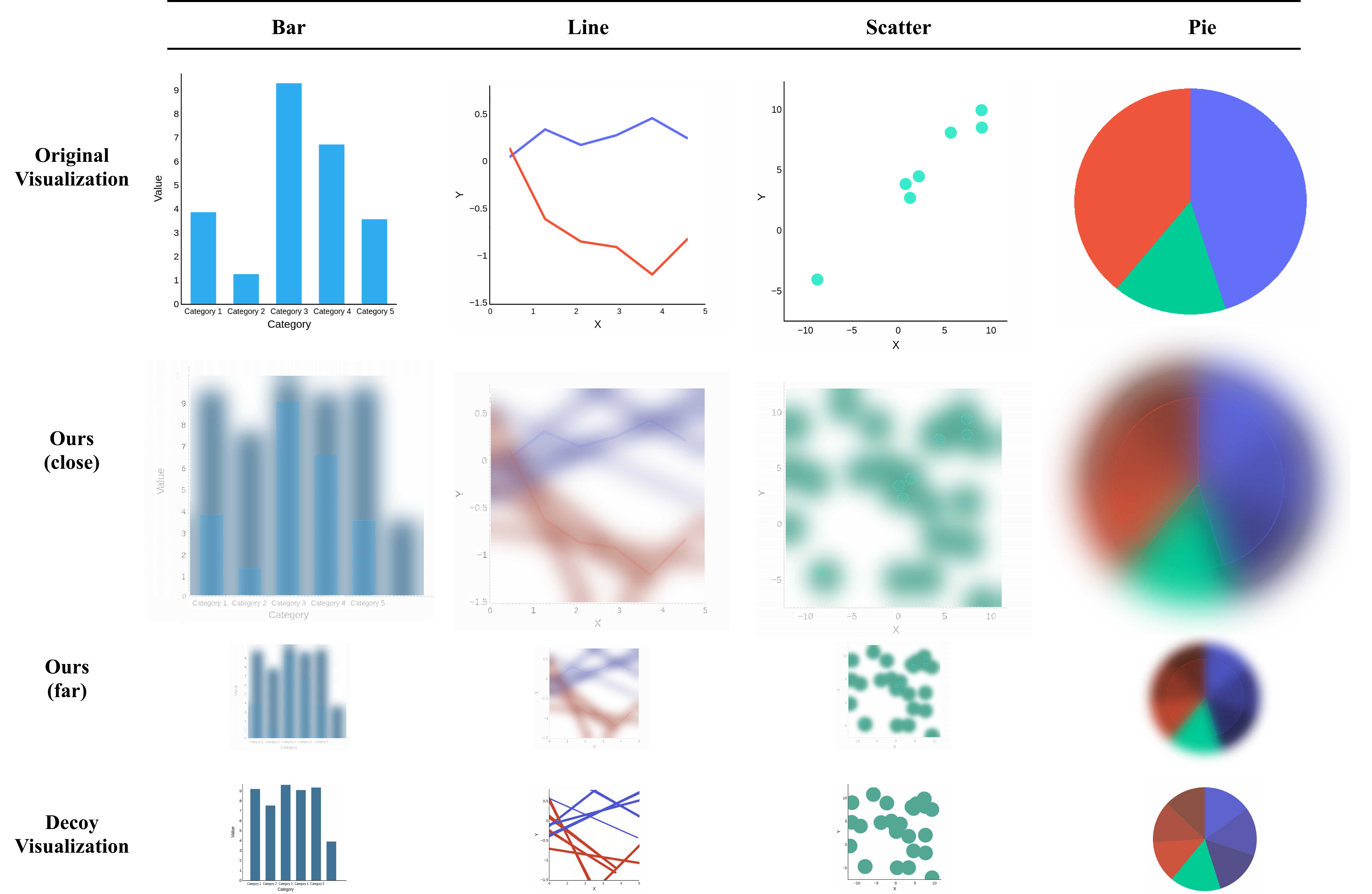}
    \caption{Examples of privacy-preserving visualizations produced by~\name{}. The figure showcases the simulated perception of original visualizations at a close distance (top row), privacy-preserving visualizations at a close distance (second row), privacy-preserving visualizations at a far distance (third row), and decoy visualizations at a far distance (bottom row). \textbf{Note}: the effect of privacy-preserving visualizations (second row) depends on their displayed sizes, and \textbf{readers are recommended to expand them to validate their effectiveness if they are displayed in a small size}.
    }
    \label{fig: case_study}
\end{figure*}

In this section, we showcase our privacy-preserving visualizations from two perspectives: (1) the effect at a close viewing distance and (2) the effect at a far viewing distance. For each perspective, we illustrate the privacy-preserving visualizations for the bar chart, line chart, scatter plot, and pie chart, respectively, as shown in Figure~\ref{fig: case_study}.

\textbf{Close viewing.} When observed at a close distance, our privacy-preserving visualizations allow users to identify the original visualization easily, as shown in the first and second rows of Figure~\ref{fig: case_study}. In the privacy-preserving bar chart, we remove the low frequencies in the original bars and high frequencies in the decoy bars, resulting in a blurry appearance of the decoy bars while the original bars remain distinct. Similarly, in the privacy-preserving line chart, the original lines consist of discretized dots, while the decoy lines appear blurry and exhibit different colors, making them easily distinguishable at a close distance. In the privacy-preserving scatter plot, the larger decoy dots serve as guides for users to locate the original dots quickly. Lastly, in the privacy-preserving pie chart, there is a slight gradient color change in the original slice. However, the solid borders of the original slices are much more visible at a close distance. This visibility encourages the perception of the enclosed area as a complete entity. As a result, it reinforces the user's ability to discern the original pie chart from the privacy-preserving pie chart.

\textbf{Far viewing.} By comparing the third and fourth rows of Figure~\ref{fig: case_study}, it is easy to observe that our privacy-preserving visualizations effectively obscure the data of the original visualization through a decoy visualization. Shoulder surfers are more inclined to perceive the decoy visualization at a far distance when observing our privacy-preserving visualizations in all four visualization types. For instance, our decoy bar chart incorporates additional bars of varying heights compared with the original bar chart. In the privacy-preserving line chart, the decoy line chart contains a greater number of lines with different trends compared to the original lines. With regard to the scatter plot, the decoy dots in the decoy scatter plot are larger than the original dots, leading to the larger decoy dots overlapping and successfully concealing the original dots in the privacy-preserving scatter plot, when shoulder surfers view the privacy-preserving scatter plot from a far distance. For the pie chart, the decoy pie has more slices than the original pie, and the split slices in the decoy pie exhibit slightly varying colors. Consequently, when shoulder surfers view the privacy-preserving pie chart, they mistakenly perceive the pie chart as having seven slices, whereas the original pie chart only has three slices.  
The privacy protection ability is attributed to the similar colors for the original and decoy visualizations, making shoulder surfers hardly differentiate them. On the other hand, the \newalexin{underlying data} between them is distinct, impeding shoulder surfers from observing the original visualization from a far viewing distance.
    
\begin{figure}[ht!]
    \centering
    \includegraphics[width=0.8\linewidth]{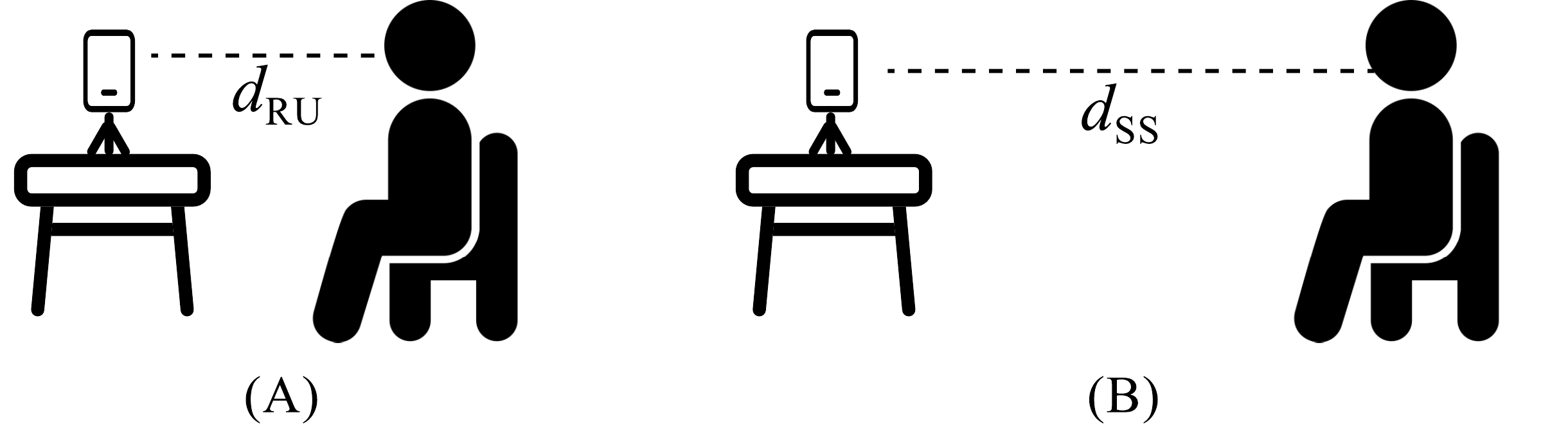}
    \caption{An illustration of our experimental setup. (A) The participant, acting as a Real User (RU), is required to view the test visualization at a distance of 30 cm (i.e., \textit{d}$_{\text{RU}}$). (B) The participant, acting as a Shoulder Surfer (SS), is required to view the test visualization at a distance of 90 cm (i.e., \textit{d}$_{\text{SS}}$).
    }
    \label{fig: experiment_setting}
    \vspace{-1em}

\end{figure}

\section{User Studies}
\red{We conducted two user studies to extensively evaluate the effectiveness and usability of \name{} in both a controlled lab environment and a real-world context.
}

\subsection{Study 1: Controlled Laboratory Experiment}

\red{We conducted a controlled laboratory experiment to evaluate the effectiveness of \name{} from two aspects: (1) a real user's ability to read the original visualization at a close distance, and (2) the likelihood of the decoy visualization misleading a shoulder surfer at a far distance.}

\subsubsection{Participants and Apparatus}
We recruited a diverse group of 32 participants from a local university. All participants had normal or corrected-to-normal vision and no color blindness, which ensured all participants had similar vision capability allowing for a fair assessment of the privacy-preserving visualization.
\revised{Our participants comprised 10 women and 22 men aged from 23 to 34}. 

The study took place in an indoor space with normal lighting conditions and lasted for 30 minutes. Before the evaluation, all the participants read and signed an IRB-approved consent form, and they were compensated with \$10 for their time and effort. To provide a realistic user experience on mobile data visualization, in the study, we used a 6.67-inch mobile phone with a resolution of 1080 x 2400 pixels. \newalexin{The phone was placed on an adjustable stand, allowing the participants to see the visualizations displayed on the mobile phone horizontally~\cite{Bce2022PrivacyScoutAV}} and comfortably, as shown in Figure~\ref{fig: experiment_setting}.

\subsubsection{Stimulus}
In our evaluation, we employed the four types of widely-used visualization: pie charts, bar charts, line charts, and scatter plots~\cite{battle2018beagle}. The original visualizations for the study were created using Plotly~\cite{plotly} with randomly generated data. 
These privacy-preserving visualizations are used as the stimulus in our user study to evaluate the effectiveness of \name{}.

We compared \name{} with two baseline methods to objectively evaluate its effectiveness:
\begin{itemize}
\item \underline{U}\alexin{n}processed \underline{V}isualization (\textit{UV}): the test visualizations without any privacy-preserving processing.
\item \underline{M}asking \underline{S}cheme-based privacy-preserving (\textit{MS}): visualizations processed by the masking scheme only~\cite{zhang2023don} (\textit{MS}). 
\end{itemize}
\alexin{Since the previous work~\cite{zhang2023don} 
 developed a method that adaptively and effectively protects the privacy of textual elements in visualizations, such as labels and values, we directly apply the approach to the textual element
 and focused solely on visualization marks in the evaluation process.}
To mitigate order effects, we randomized the presentation of 48 test visualizations (4 visualization types × 4 instances × 3 methods) for each participant, ensuring a thorough evaluation of \name{} compared to the baselines.

\subsubsection{Procedure}

The close viewing distance was set at 30cm between the participants and the mobile phone (Figure~\ref{fig: experiment_setting} (A)), reflecting a typical distance when a person uses a mobile phone~\cite{yoshimura2017smartphone}. Additionally, the far distance was set at 90cm (Figure~\ref{fig: experiment_setting} (B)), simulating a scenario where a shoulder surfer sits behind a user's seat in public transportation, such as buses\alexin{~\cite{SeatWidth2021}}.

To emulate shoulder surfing attacks and user experiences, we divided the 32 participants into two groups: Shoulder Surfers (SSs) and Real Users (RUs). The SS participants, unaware of the methods applied to the test visualizations, were instructed to view the visualizations from a distance. \red{Prior research has shown that typical shoulder surfing attacks involve average gaze durations of 2.1 seconds and maximum durations of 7.53 seconds~\cite{abdrabou2022understanding}, while glanceable visualizations can be comprehended within 2 seconds~\cite{lee2021mobile}. To ensure a certain degree of fault tolerance, all of the} participants were given up to 5 seconds to identify the original visualizations, simulating a brief glance at a victim's screen. The RU participants were familiar with the mechanism of the three methods and were asked to identify the original data from the test visualizations. To maintain the objectivity of the study, the RUs were not informed which method was ours during the evaluation process.

After viewing each visualization in the study, participants would select a visualization that most closely matched what they saw from four different visualizations provided as choices, where a test example is shown in Appendix A.
Besides the correct choice, the other three choices are of the same type as the correct one.

\subsubsection{Overall Recognition Accuracy}
\label{expm:result}

Our primary outcome measure is recognition accuracy~\cite{chen2019keep}, which captures how often participants can correctly identify the protected information (i.e., the original visualization) from the test visualizations (i.e., privacy-preserving visualizations). Based on the roles of the participants, we distinguish between shoulder surfer recognition accuracy and user recognition accuracy. Shoulder surfer recognition accuracy reflects the privacy protection capability of a method against shoulder surfing attacks, where lower values are better. In contrast, user recognition accuracy reflects how well a method preserves the readability of the original visualizations for legitimate users, where higher values are better. Since normality tests indicated that the data followed a non-normal distribution, we applied the Wilcoxon signed-rank test to examine pairwise differences among the methods~\cite{Wang2018AVF}.

\begin{figure}[ht!] 
\centering \includegraphics[width=\linewidth]{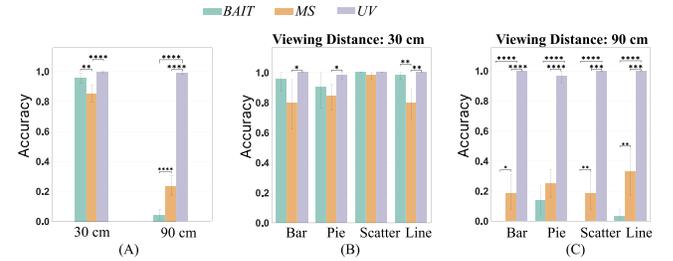} \caption{The user recognition accuracy and shoulder surfer recognition accuracy in the user study, accompanied by their 95\% Confidence Intervals (CI). \name{}, \textit{MS}, and \textit{UV} denote the test visualizations processed using \name{}, masking scheme, and without any processing, respectively. (A) The user recognition accuracy (at 30 cm) and shoulder surfers recognition accuracy (at 90 cm) for the original visualization in the test visualizations, respectively. (B) The user recognition accuracy (at 30 cm) for original visualization in test visualizations across four visualization types. (C) The shoulder surfer recognition accuracy (at 90 cm) for original visualization content in test visualizations across four visualization types. 
Friedman tests confirmed significant overall differences among the three methods at both viewing distances (\textit{p} $<$ 0.001). 
* indicates a statistically significant difference with \textit{p} $<$ 0.05, ** indicates \textit{p} $<$ 0.01, *** indicates \textit{p} $<$ 0.001, and **** indicates \textit{p} $<$ 0.0001.} \label{fig: study_rating} 
\end{figure}

Figure~\ref{fig: study_rating} provides an overview of how \name{}, \textit{MS}, and \textit{UV} perform for both Real Users (RUs) and Shoulder Surfers (SSs). Figure~\ref{fig: study_rating}~\component{A} aggregates recognition accuracy across all visualization types. At the close viewing distance of 30~cm, \name{} achieves high user recognition accuracy that is very close to the accuracy of unprocessed visualizations, indicating that RUs can read \name{}-protected visualizations nearly as easily as unprocessed ones. In contrast, \textit{MS} shows noticeably lower accuracy, suggesting a loss of readability for the masking scheme.

\revised{At the far viewing distance of 90~cm (Figure~\ref{fig: study_rating}~\component{A}), \name{} reduces the shoulder surfer recognition accuracy to nearly 0\%, demonstrating strong privacy preservation. In contrast, \textit{UV} maintains a recognition accuracy close to 100\%, meaning that shoulder surfers can almost always correctly identify the original visualization from unprocessed visualizations. \textit{MS} achieves moderate accuracy between the two, providing only partial privacy protection. Notably, since participants selected from four choices, the random guessing baseline is 25\%. While \textit{MS} reduces accuracy to near-random levels, \name{} pushes accuracy well \textbf{below} this baseline, indicating that shoulder surfers are not merely failing to identify the original visualization, but are actively being \textbf{misled} by the decoy visualization into selecting incorrect options.}

\revised{To provide a direct quantitative comparison, we report the overall mean recognition accuracy and standard deviation for \name{} and \textit{UV} at both viewing distances. At 30~cm, \name{} achieved a mean accuracy of 0.96 (SD = 0.09), compared to 0.99 (SD = 0.02) for \textit{UV}. At 90~cm, \name{} reduced accuracy to 0.04 (SD = 0.09), whereas \textit{UV} remained at 0.99 (SD = 0.03). A Wilcoxon signed-rank test confirmed that the difference between \name{} and \textit{UV} at 90~cm is statistically significant (\textit{p} $<$ 0.0001), demonstrating that \name{} provides substantially stronger privacy protection than unprocessed visualizations while maintaining comparable readability at close viewing distances.}

\revised{Figure~\ref{fig: study_rating}~\component{B} and~\component{C} further break down recognition accuracy by visualization type. At 30~cm (Figure~\ref{fig: study_rating}~\component{B}), \name{} maintains high user recognition accuracy across bar, pie, scatter, and line charts, very close to that of unprocessed visualizations (\textit{UV}), whereas \textit{MS} shows noticeable accuracy drops for some chart types (e.g., bar and line), suggesting that \textit{MS} harms readability. At 90~cm (Figure~\ref{fig: study_rating}~\component{C}), \name{} reduces shoulder surfer recognition accuracy to nearly 0\% for bar and scatter plots, and achieves substantially lower accuracy than \textit{UV} for all chart types, demonstrating that \name{} provides effective privacy protection across different visualization types.}

\revised{In summary, Figure~\ref{fig: study_rating} demonstrates the overall effectiveness of \name{}: at the close viewing distance (30~cm), \name{} preserves high user recognition accuracy comparable to unprocessed visualizations; at the far viewing distance (90~cm), \name{} reduces shoulder surfer recognition accuracy to near-random levels, providing strong privacy protection. As one RU summarized, ``We can clearly distinguish the original visualization from the decoy visualization; the only price is a little extra time and effort to adapt.'' In the following, we introduce the statistical procedures that we use to quantify these observations and then report the detailed quantitative results.}

\subsubsection{Detailed Pairwise Comparison Across Viewing Distances and Chart Types}
To further identify the detailed differences among the three methods, we conducted pairwise comparisons of each pair of methods across  \textit{different viewing distances} and \textit{different chart types}.


\begin{table}[!htb] \centering \caption{\revised{Effect sizes in terms of absolute value between three methods at different viewing distances. The negligible effect size is highlighted.}} \begin{tblr}{ hlines, vlines, } \diagbox{Method Comparison}{Viewing Distance} & 30 cm & 90 cm \\ \textit{MS} v.s. \textit{\name{}} & 0.26660 & 0.46167 \\ \textit{MS} v.s. \textit{UV} & 0.34595 & 0.96582 \\ \textit{\name{}} v.s. \textit{UV} & \textbf{0.07861} & 1.00 \end{tblr} \label{table: general_effect_size} \end{table}

\textbf{Pairwise Differences Across Viewing Distances.}
\alexin{Table~\ref{table: general_effect_size} \revised{shows the effect sizes, represented as Cliff's Delta values, for three distinct methods (\textit{MS}, \name{}, and \textit{UV}) at two viewing distances: 30 cm and 90 cm. Cliff's Delta is a non-parametric metric that quantifies the magnitude of differences between two data sets. Specifically, Cliff's Delta absolute value ranges from 0 to 1, and classifies the value below 0.147 as negligible, between 0.147 and 0.33 as small, 0.33 to 0.474 as medium, and anything above 0.474 as a large effect size~\cite{cliff2014ordinal}. A ``negligible" effect size indicates that the differences between methods are minimal. In contrast, a large effect size highlights substantial differences. For the short viewing distance of 30cm, the difference between \name{} and \textit{UV} (0.07861) is negligible, whereas the difference between \textit{MS} and \textit{UV} (0.34595) presents as a medium effect size. These findings suggest that \name{} allows visualization owners at proximity to see and analyze the original data with a comparable performance as unprocessed visualizations (\textit{UV}), outperforming the masking scheme-based method (\textit{MS}). For the longer viewing distance of 90 cm, the large effect size between \name{} and \textit{UV} (1.00) indicating that \name{} provides substantially stronger privacy protection than unprocessed visualizations.}}

\textbf{Pairwise Differences Across Visualization Types.}
\revised{Table~\ref{table: close_effect_size} provides detailed statistics of the effect sizes for method comparisons at the close viewing distance (30 cm). For this distance, the differences between \name{} and \textit{UV} across \textbf{all four visualization types} are negligible, further emphasizing \name{}'s ability to \textbf{retain the visibility} of original visualizations at this distance. In contrast, the comparison between \textit{MS} and \textit{UV} on bar chart, line chart and pie chart shows small to medium effects, indicating that the masking scheme introduces noticeable distortions for some chart types.}

\newalexin{Qualitative comments from RUs echoed these trends: participants frequently reported that \name{}-preserved bar and line charts were “as easy to read as normal”, whereas MS-processed charts were described as “slightly harder to distinguish” in dense regions.}


\begin{table*}[!htb]
\centering
\caption{
The table depicts effect sizes in terms of absolute value between three methods across four visualization types at the close viewing distance (30 cm). The negligible effect size is highlighted.}
\begin{tabular}{|l|l|l|l|l|} 
\hline
\diagbox{Vis. Type}{Method Comparison} & Bar     & Line    & Scatter & Pie      \\ 
\hline
\textit{MS }v.s. \textit{\name{}}         & 0.21094 & 0.51172 & \textbf{0.06250} & 0.29688  \\ 
\hline
\textit{MS }v.s. \textit{UV}           & 0.31250 & 0.56250 & \textbf{0.06250} & 0.44531  \\ 
\hline
\textit{\name{} }v.s. \textit{UV}         & \textbf{0.12500} & \textbf{0.06250} & \textbf{0.00000} & \textbf{0.12891}  \\
\hline
\end{tabular}
\label{table: close_effect_size}
\end{table*}

\begin{table*}[!htb]
\centering
\caption{
The table depicts effect sizes in terms of absolute value between three methods across four visualization types at the far viewing distance (90 cm). The large effect size is highlighted.}
\begin{tabular}{|l|l|l|l|l|} 
\hline
\diagbox{Vis. Type}{Method Comparison} & Bar              & Line    & Scatter & Pie      \\ 
\hline
\textit{MS }v.s. \textit{\name{}}         & 0.43750          & \textbf{0.55469} & \textbf{0.50000} & 0.31641  \\ 
\hline
\textit{MS }v.s. \textit{UV}           & \textbf{1.00000}          & \textbf{0.93750} & \textbf{0.93750} & \textbf{0.99219}  \\ 
\hline
\textit{\name{} }v.s. \textit{UV}         & \textbf{1.00000} & \textbf{1.00000} & \textbf{1.00000} & \textbf{1.00000}  \\
\hline
\end{tabular}
\label{table: far_effect_size}
\end{table*}

\revised{Table~\ref{table: far_effect_size} summarizes the corresponding effect sizes at 90 cm across the four visualization types. The Cliff's Delta values indicate large effects for \name{} across all four visualization types, matching the \textbf{substantial reductions} in shoulder surfer recognition accuracy compared with \textit{UV}. Participants acting as SSs experienced considerable difficulty in identifying original visualizations processed by \name{}, with recognition accuracy dropping to near-random levels for certain visualization types. This justifies the effectiveness of \name{} in providing privacy protection across popular visualization types. SSs' open-ended feedback supported these quantitative findings: many reported that \name{}-protected bar and scatter plots at 90 cm ``looked like noise'' or that they could ``only guess'' the underlying trends, whereas unprocessed charts were described as ``immediately readable''.}

\subsubsection{Summary of Findings}
\revised{In summary, Study~1 demonstrates that \name{} effectively balances privacy protection and readability. The key findings are:}

\begin{itemize}
    \item \textbf{High readability at close proximity.} 
    At the close viewing distance (30~cm), \name{} achieves user recognition accuracy that is perceptually comparable to unprocessed visualizations (\textit{UV}). Legitimate users can easily identify the original data, suggesting that the overlaid privacy protection does not hinder real users.
    
    \item \textbf{Robust privacy protection at a distance.} 
    At the far viewing distance (90~cm), \name{} consistently reduces shoulder surfer recognition accuracy to near 0\%, significantly below the random guessing baseline (25\%). Unlike the masking scheme (\textit{MS}) which merely blurs content, \name{} actively misleads shoulder surfers into perceiving the decoy visualization.
    
    \item \textbf{Statistical validation via effect size analysis.} 
    We statistically confirmed the above observations with detailed pairwise comparisons (Cliff's Delta) for four visualization types and two distances. The analysis reveals a distinct pattern: \textbf{negligible effect sizes} at the close distance (confirming no significant loss of readability) versus \textbf{large effect sizes} at the far distance (confirming substantial privacy protection). This statistical evidence proves that our decoy visualizations are robust across different chart types.
\end{itemize}

\subsection{Study 2: Scenario-based Usability Evaluation}

\red{While Study 1 established the effectiveness of \name{} in controlled laboratory conditions, it is equally important to further understand how \name{} performs in real-world contexts where users typically interact with visualizations. This second study was designed to evaluate the usability and practical value of \name{} in real-world environments where privacy concerns are particularly relevant.}

\subsubsection{Participants and Settings}

\red{To maintain the consistency and objectivity of the research, we recruited the 12 participants (P1-P12), who had been involved in our formative interviews (Section 3), for an in-depth study.}

\red{The study was conducted in real-world environments where visualization privacy is particularly relevant, including busy café and university corridor. These locations were selected to simulate common scenarios where mobile visualizations are used while privacy threats are present. Each session lasted approximately 30 minutes.}

\subsubsection{Procedure}

\red{The study followed a four-phase plan:}

\begin{enumerate}

\item \textbf{Scenario Immersion:} \red{Participants were first engaged in a scenario immersion exercise where they were asked to imagine themselves in situations requiring privacy protection while viewing data visualizations. Example scenarios included reviewing financial performance data visualizations in a public cafe or analyzing confidential research results.}

\item \textbf{Prototype Interaction:} \red{Participants were presented with pairs of visualizations, original versions and their corresponding \name{}, processed versions, across different visualization types. The visualizations incorporated synthetic data aligned with the scenarios explored during the immersion phase. Participants engaged in tasks designed to emulate authentic analytical workflows, such as: (1) Identifying trends in time-series data using line charts, (2) Discovering clusters and patterns in scatter plots, (3) Comparing proportions in pie charts, and (4) Identifying maximum and minimum values in bar charts.
\red{Participants were free to choose tasks that aligned with their expertise and interests.}
\item \textbf{Semi-structured Interviews:} Following the interaction phase, participants engaged in semi-structured interviews focused on three key areas:
\begin{enumerate}
\item \textbf{Usability:} What challenges did participants encounter when using \name{}-processed visualizations for analytical tasks?
\item \textbf{Real-world application value:} How valuable is \name{} for protecting visualization privacy in real-world contexts?
\item \textbf{Comparative assessment:} How does \name{} compare to traditional privacy protection approaches such as physical privacy screens or behavioral practices (e.g., manually shielding screens with hands)?
\end{enumerate}}

\red{\item \textbf{Data Collection:} 
In this study,
we collected both qualitative data (verbal feedback and interview responses) and quantitative data (NASA-TLX~\cite{hart1988development} and PSSUQ~\cite{lewis1995ibm} questionnaire responses),
which were later analyzed using thematic analysis to identify patterns in usability challenges, perceived benefits, and contextual factors.}
\end{enumerate}

\subsubsection{Results}

\revised{We report the results of Study~2 in terms of the usability, workload, learnability and adaptation, perceived practical value of \name{} in real-world settings, and its comparison with existing approaches.}

\begin{figure*}[ht!]
    \centering
    \includegraphics[width=\linewidth]{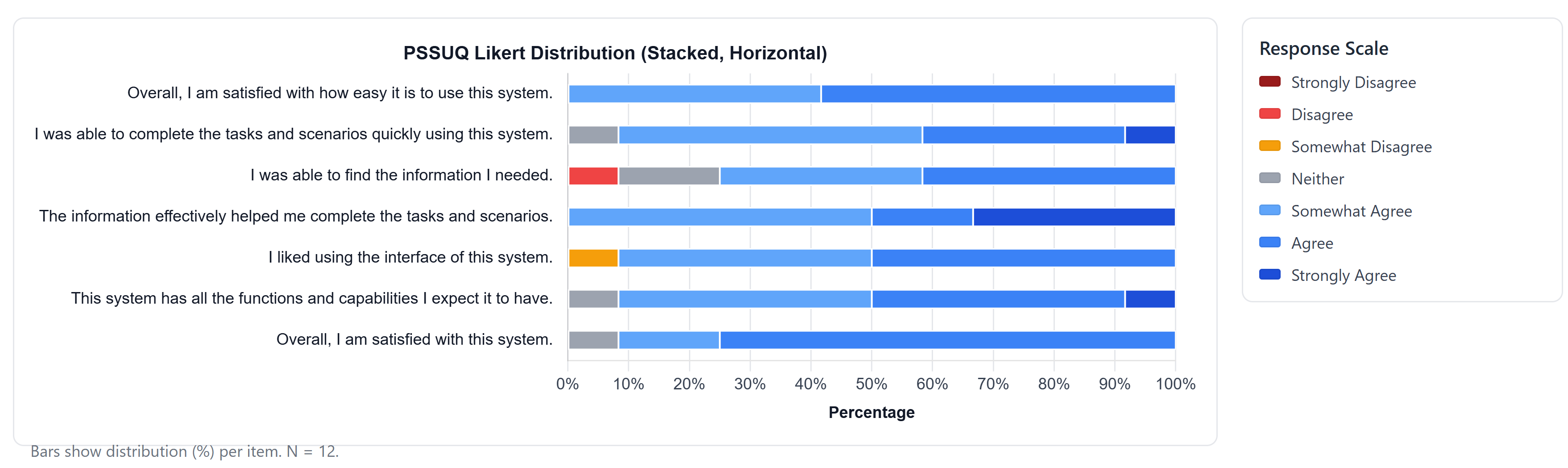}
    \caption{Usability and satisfaction results for \name{} in Study~2, as measured by selected PSSUQ items. Higher values indicate better perceived usability.}
    \label{fig: study2_pssuq}
\end{figure*}

\textbf{Usability}.
\revised{We used the Post-Study System Usability Questionnaire (PSSUQ)~\cite{lewis1995ibm} to assess perceived usability and satisfaction with \name{}. The PSSUQ is a widely used, technology-agnostic questionnaire developed at IBM for post-task usability evaluation. It consists of 16 items that yield an overall usability score and three subscales---System Usefulness, Information Quality, and Interface Quality. All items are rated on a seven-point Likert scale, with lower scores indicating better perceived usability. Prior work has shown that the PSSUQ exhibits high internal reliability and sensitivity to differences between interfaces, and that it can provide informative results even with relatively small sample sizes.}
\revised{Because \name{} is implemented as a visualization ``plugin'' rather than a complete interactive system, we adopted only those PSSUQ items that were directly relevant to our setting. In particular, we focused on three aspects: (1) how easy \name{} is to use, (2) whether users can find the information they need in \name{}-processed visualizations, and (3) how satisfied users are with \name{} overall. The item-level response distributions for these questions are shown in Figure~\ref{fig: study2_pssuq}.}
\revised{Overall, participants responded positively across all seven PSSUQ items. Regarding \textbf{ease of use}, 11 out of 12 participants agreed that \name{} was straightforward to learn and enabled efficient task completion. In terms of \textbf{information quality}, a strong majority (at least 9 out of 12) indicated that they could successfully locate the necessary information and that the data provided was effective for solving the scenarios. Feedback on \textbf{interface quality} was similarly positive, with 11 participants expressing satisfaction with the interface design. Finally, regarding \textbf{overall satisfaction}, 11 out of 12 participants confirmed that the system met their functional expectations and provided a satisfactory user experience, suggesting that \name{} is effective as a privacy-preserving visualization solution.}
\revised{In short, the feedback from participants confirms that \name{} is easy to use, provides sufficient information for typical analytic tasks, and enables effective privacy protection, which is well-aligned with the results of Study~1.}

\begin{table}[!htb]
\centering
\caption{Summary of NASA-TLX scores for viewing \name{}-processed visualizations in Study~2 (n = 12). Values are means with standard deviations in parentheses.}
\begin{tabular}{lcc}
\hline
\textbf{Dimension} & \textbf{Mean} & \textbf{SD} \\
\hline
Mental Demand      & 35.83 & 23.53 \\
Physical Demand    & 20.42 & 17.12 \\
Temporal Demand    & 39.58 & 13.22 \\
Performance        & 16.67 & 9.85  \\
Effort             & 40.42 & 16.16 \\
Frustration        & 10.42 & 6.20  \\
\hline
Overall TLX (0--100) & 27.22 & 11.13 \\
\hline
\end{tabular}
\label{table:study2_nasatlx}
\end{table}

\textbf{Workload}.
\revised{
We used the NASA Task Load Index (NASA-TLX)~\cite{hart1988development} to estimate the subjective workload of viewing \name{}-processed visualizations in realistic scenarios. NASA-TLX conceptualizes \emph{mental workload} as the overall perceived cost in cognitive, perceptual, and emotional resources required to carry out a task, and decomposes it into six dimensions: mental, physical, and temporal demand, perceived performance, effort, and frustration. Table~\ref{table:study2_nasatlx} summarizes the mean scores and standard deviations for each dimension on a 0--100 scale, as well as the overall raw workload score.}
\revised{
Prior work has shown that the original weighting procedure in NASA-TLX, which requires participants to perform pairwise comparisons among dimensions and derive individualized weights, adds considerable complexity and time burden while providing little improvement in the psychometric quality of the scale~\cite{nygren1991psychometric}. Following these recommendations, we adopt the Raw TLX scoring approach and compute the overall workload score as the arithmetic mean of the six dimension ratings.}
\revised{Overall, \name{} received an average raw TLX score of 27.22/100 (SD = 11.13), which corresponds to a relatively low level of subjective workload. Among individual dimensions, mental demand (M = 35.83, SD = 23.53) and effort (M = 40.42, SD = 16.16) both fall well below the mid-point of the scale, indicating that participants experienced some cognitive engagement when interpreting \name{}-processed charts but did not find the tasks mentally taxing. It shows that the additional visual structure introduced by \name{} (e.g., overlaid decoy patterns and frequency adjustments) imposes an acceptable mental burden. Temporal demand is also moderate (M = 39.58, SD = 13.22), suggesting that participants did not feel rushed when completing the tasks in realistic public settings.}
\revised{In addition, physical demand (M = 20.42, SD = 17.12) and frustration (M = 10.42, SD = 6.20) are both very low. The low frustration score in particular indicates that \name{} does not induce substantial stress, annoyance, or discouragement during use, even though it alters the appearance of standard visualizations. For the performance dimension, we followed the common practice that lower values mean better perceived performance. The resulting low mean performance score (M = 16.67, SD = 9.85) therefore indicates that participants felt they performed the tasks successfully and were satisfied with \name{}-processed visualizations. This subjective sense of high task success is consistent with the objective recognition results reported in Study~1, where real users achieved high accuracy when interpreting \name{}-protected charts at close viewing distances.}
\revised{In summary, \name{} requires only modest additional cognitive resources beyond those needed for reading standard visualizations, yet does not make users feel stressed or overloaded.}

\textbf{Learnability and Adaptation}.
Despite the overall positive feedback on workload and usability, Figure~\ref{fig: study2_pssuq} shows that two participants (P3 and P8) expressed some reservations about the usability of our approach, as indicated by lower ratings on selected PSSUQ items. Follow-up interviews revealed that these concerns were primarily related to an initial adaptation period: participants needed time to get used to the visualizations processed by \name{}, which contain an overlaid decoy visualization in the background and adjusted spatial frequencies.
Both P3 and P8 reported that, at first, interpreting \name{}-processed visualizations required more visual effort. However, they emphasized that this additional burden diminished after a short period of familiarization, and that their efficiency in interpreting the protected visualizations improved notably over time. Motivated by this observation, we also conducted follow-up inquiries with other participants to probe the learning curve of \name{} more systematically. Their feedback echoed that of P3 and P8: participants described a smooth learning curve and reported that they quickly became accustomed to the appearance and behavior of \name{}-protected charts.
These qualitative findings complement the NASA-TLX results: although a brief adaptation period is needed to understand how \name{} modifies standard visualizations, the ongoing mental workload and frustration remain low once users are familiar with the design.

\textbf{Practical Application Value}.
Participants broadly endorsed the real-world usefulness of \name{} for automatically generating privacy-preserving visualizations that can be viewed in public or semi-public settings. They perceived \name{} as a viable option in scenarios where they did not want the sensitive information encoded in a visualization to be visible to others, such as viewing monthly income trends in a banking app or monitoring decreasing daily exercise time in a fitness dashboard.
Beyond such everyday examples, P1, a postgraduate student, pointed out that \name{} could also help protect the confidentiality of preliminary experiment results when checking charts on a university bus. In summary, whenever the underlying data represented by a visualization is sensitive, participants felt that \name{} provides a practical additional layer of protection without fully sacrificing readability at close distances.

\textbf{Comparison with Existing Privacy-Preserving Approaches}.
When asked to compare \name{} with other privacy-preserving approaches, participants mainly focused on hardware-based privacy screens. While acknowledging the effectiveness of privacy screens in narrowing the \emph{viewing angle}, several participants (e.g., P6 and P9) reported frustration with their inherent drawbacks, such as added cost, the need to purchase and attach accessories, and reduced display brightness.
By contrast, \name{} was praised for its software-based design, which participants found more convenient and flexible. They particularly appreciated what they described as the ``zero-cost and integrated'' nature of the solution. As P2 remarked, ``This is better because I don't have to go out and buy another accessory for my phone.'' These comparative comments underscore a key advantage of \name{}: it provides on-demand, situational privacy without requiring physical accessories or incurring additional cost, making it easier to adopt in day-to-day usage.



\subsubsection{Summary of Findings}
\revised{Study~2 confirms that \name{} is not only effective but also usable and practical in realistic environments. The key findings, synthesized from both quantitative metrics and qualitative feedback, are as follows:}

\begin{itemize}
    \item \textbf{High usability and low perceived workload.} 
    Quantitative results from the PSSUQ indicate high satisfaction with system usefulness and information quality, while NASA-TLX scores reveal a low overall workload. This evidence suggests that the added layer of privacy protection does not impose a significant cognitive burden or frustration on users in real-world scenarios.
    
    \item \textbf{Manageable adaptation curve.} 
    Qualitative interviews clarify that while the unique visual design of \name{} requires a brief period of familiarization, the learning curve is smooth. Participants reported that after a short adaptation phase, they could efficiently interpret the visualizations.
    
    \item \textbf{Effectiveness in practical applications.} 
    Our user study in cafes and corridors confirmed that \name{} remains effective outside the lab. Participants validated its practical value as a flexible, software-based solution that provides ``on-demand'' privacy. Rather than completely replacing hardware solutions, \name{} serves as a convenient and cost-free alternative to physical privacy screens, particularly suitable for dynamic public environments.
\end{itemize}

\section{Discussion}


\subsection{Design Implications and Lessons Learned}

\revised{In this section, we shared the lessons learned from our study and the corresponding implications, including the inherent privacy protection challenge of data visualization that originates from its glanceability, one-time cognitive and learning cost of perception-driven privacy-protection method, and visual-illusion-informed privacy protection.}



\revised{
\textbf{{Visualization Privacy} vs. {Visualization Glanceability}}.
There has been much research work on data privacy protection, especially for text data or numerical data~\cite{andow2017uiref,huang2015supor}.
Different from texts or numerical data, which requires sequential cognitive processing, data visualizations are inherently glanceable. 
Visualizations leverage preattentive visual attributes to communicate data insights almost instantaneously~\cite{lee2018data,Kerren2008Information,few2013information}.
This high glanceability enables efficient information processing and is a key strength of data visualizations. However, this efficiency also introduces a unique privacy challenge. The same mechanisms that enable legitimate users to quickly extract insights also allow unauthorized observers to rapidly acquire sensitive information through a brief and casual glance.
\name{}'s design addresses this challenge by introducing a seemingly plausible yet misleading decoy visualization. It offers an effective approach for privacy protection of glanceable visualizations, which cannot be achieved by existing privacy protection methods designed for texts or numerical data.
}

\revised{
\textbf{{One-time Cognitive and Learning Cost} vs. {Effective Privacy Protection}}.
As shown in our above evaluations, the privacy-preserving visualizations processed by \name{} needs one-time cognitive and learning cost from users. Specifically, users need to know what is the actual visualization and the decoy visualization. 
Our user study findings suggest that this cognitive cost is adaptive and diminishing. Participants reported that the initial unfamiliarity rapidly decreased with a brief interaction with the visualizations processed by \name{}.
We speculate that participants' quick learning and adaption to our approach is due to the intrinsic simplicity and intuitiveness of the working mechanism of \name{}, i.e., the one-time cognitive and learning cost is minimum.
We recommend that such a practice of minimum one-time cognitive and learning cost should also be followed in the development of future perception-driven privacy protection methods.
}

\revised{
\textbf{{Visual Illusions for Visualization Privacy Protection}}.
The core innovation of \name{} was inspired by prior research on visual illusions, an important research direction in the fields of perception and psychology.
Specifically, our approach is inspired by the principles of Hybrid Images~\cite{oliva2006hybrid}. By exploiting the distinct spatial frequency processing channels of the human visual system, we create a divergence in perception based on viewing distance. Our work confirms that privacy protection can be achieved through perceptual manipulation built upon existing research on visual illusion.
This work suggests a promising but still underexplored research avenue, i.e., systematically mining
visual
illusions to construct new privacy protection methods.
For example, 
it may be possible to leverage the Crowding Effect~\cite{levi2008crowding}, where target objects become unidentifiable when surrounded by specific patterns in peripheral vision, to hide sensitive visual information. 
}

\subsection{Limitations and Future Work}
We further discuss the possible limitations of \name{} in this section, which also inform the possible future research directions.

\textbf{Generalizability to various visualization types and devices.}
In this study, we applied \name{} to four popular visualization types due to their popularity on the internet~\cite{battle2018beagle}. It is crucial to emphasize that \name{}'s applicability is not limited to these four visualization types. The core strategy behind our approach is the generation of a decoy visualization designed to distract shoulder surfers. This decoy visualization takes into account both data distribution and visual properties of the original visualization. Given that all visualizations include these features, \name{} has great potential to be extended to other types of visualizations such as graph visualizations and even infographics. 
Moreover, \name{} is not limited to a specific device. Although our evaluation employed only one device to control experimental variables, the method is designed to be adaptable across different devices.
By providing the visualization image size shown on mobile devices, \name{} can automatically generate the privacy-preserving visualization for the corresponding original visualization, balancing good visibility at proximity and privacy preservation beyond a certain viewing distance.
In future work, we plan to extend \name{} to other types of visualizations and different mobile devices.

\textbf{Extending privacy-preserving techniques to animated visualizations.}
The existing research on privacy-preserving mobile data visualizations, including \name{}, focuses on
static visualizations displayed on mobile devices~\cite{zhang2023don}. However, with the increasing capability of mobile phones,
animated visualizations have become increasingly popular
on mobile phones 
due to their engaging characteristics and effectiveness in conveying complex information~\cite{brehmer2019comparative, zeng2023semi}.
Despite their intriguing format and capability of conveying the underlying data, animated visualizations are also vulnerable to privacy preservation risks including shoulder surfing attacks.
Compared to static visualizations, it is more challenging to achieve privacy preservation for animated visualizations due to their dynamic nature and dense information displayed within a short period of time.
A promising avenue for future research is to explore how \name{} can be extended to generate privacy-preserving animated visualizations.

\textbf{Sample Size and Statistical Generalizability.}
Although our evaluation demonstrated strong and consistent results, the sample sizes (32 participants for Study 1 and 12 for Study 2) are limited from a statistical perspective. While these numbers align with typical standards in HCI and visualization research, they nonetheless constrain the statistical power of our analysis. In terms of breadth, future studies with larger and more diverse samples are needed to generalize our findings across populations with varying visual or cognitive conditions. In terms of depth, longer-term field deployments could enhance statistical validity and yield richer longitudinal insights into user behavior.
\section{Related work}
In this section, we review previous research related to our work from two perspectives: privacy-preserving visualization and visible watermark.

\subsection{Privacy-preserving Visualization}

\textbf{Identity leakage: }In contemporary society, safeguarding sensitive information from potential adversaries during data transformation poses a significant challenge. As many datasets may contain sensitive information about individuals, analyzing and visualizing these datasets may result in the disclosure of personal information~\cite{Sarathy2023DontLA}. \newalexin{Various data anonymization techniques have been proposed to encrypt the identifiers that can be used to directly or indirectly identify an individual from the data points within a dataset~\cite{chou2016obfuscated,Chou2018AnES,dasgupta2012conceptualizing, Chou2018PrivacyPV, dasgupta2019guess,Chen2020FederatedVA}}. For instance, Wang~\textit{et al.}~\cite{wang2018graphprotector} combined multiple anonymization schemes to create a hybrid approach for privacy-preserving social network visualizations. This approach prevents the disclosure of real-world identities of data items from the visualization. Additionally, Dasgupta~\textit{et al.}~\cite{dasgupta2013measuring} proposed an adaptive method for privacy-preserving parallel coordinates, which clusters similar data points and bins them according to their value ranges. Similarly, Oksanen~\textit{et al.}~\cite{oksanen2015methods} proposed a privacy-preserving heatmap of trajectory data that uses three data processing techniques to prevent the disclosure of data items' identities in the heatmap.

\textbf{Shoulder surfing attacks:}
To prevent mobile visualizations from shoulder surfing attacks, a privacy film can be applied to the screen, limiting its visible range and angle. However, this hardware-based approach incurs additional expenses and reduces users' visibility of the screen~\cite{ali2014protecting}. \red{Previous software-based privacy-preserving techniques have primarily focused on protecting low-level visual elements and fine-grained details. For example, existing methods target specific textual content~\cite{papadopoulos2017illusionpin} or employ grid-based obfuscation techniques for individual text blocks and image regions~\cite{chen2019keep}. Similarly, approaches for authentication scenarios concentrate on concealing discrete interface elements such as PIN entries or individual button interactions~\cite{tang2023eye}. However, these fine-grained protection strategies are fundamentally inadequate for data visualizations, where meaningful insights can be extracted from high-level visual patterns rather than specific textual or graphical details.}

In addition to privacy films, Zhang~\textit{et al.}~\cite{zhang2023don} were the first to propose a method protecting visualization content from shoulder surfing attacks. Specifically, they considered the varying sensitivity of the Human Vision System at different viewing distances and proposed a two-stage masking scheme to transform the original visualization into a privacy-preserving one. Nevertheless, this method relies on manual parameter adjustments derived from multiple experiments, making it non-scalable and unstable under different conditions. Moreover, it demands additional user effort and time to view the visualization at close range. In contrast, leveraging \red{visual illusion} and Human Vision System, we propose a software-based approach to achieve automatic privacy-preserving visualization generation. 

\subsection{Visible Watermark}
Visible watermarking is a widely used technique to authenticate and protect digital content (e.g., images). The technique involves superimposing two images, namely a watermark (e.g., logo) and the original image~\cite{dekel2017effectiveness}. A good visible watermark should not obscure the original image details beneath it~\cite{mohanty2000dct}. To achieve this requirement, various methods have been proposed, which can be classified into spatial-based and frequency-based approaches~\cite{begum2020digital}. The spatial-based approach changes the pixel value of the original image to add the watermark. For instance, Braudaway proposed the visible watermark technique in digital images~\cite{braudaway1997protecting}. This method specifically changes the original image pixel brightness value through multiplication with the corresponding value in the watermark. The frequency-based approach applies watermark superposition in the frequency domain, which is more efficient than the spatial domain operation~\cite{meng1998embedding}. For example, Meng \textit{et al.} imposed the watermark on an original image by operating on the coefficients of the image in the discrete cosine transform (DCT) domain~\cite{meng1998embedding}. However, the approach can cause visual discontinuity~\cite{hu2001wavelet}. To overcome this limitation, a wavelet domain operation has been proposed for visible watermark generation~\cite{hu2001wavelet}. Additionally, a contrast-sensitive visible watermarking scheme has been proposed that adaptively changes the watermark intensity according to image content and Human Vision System characteristics~\cite{huang2006contrast}.

\section{Conclusion}
This paper presents \name{}, \red{a novel method that utilizes visual illusion} to automatically generate privacy-preserving data visualizations for mobile devices. \name{} also explicitly incorporates crucial factors in HVS that significantly influence human perception of data visualizations.
Specifically, \name{} overlaps a decoy visualization over a given original visualization. The resulting privacy-preserving visualization ensures that users at close proximity can perceive the original visualization content, while shoulder surfers at farther distances are redirected to the decoy visualization, which lacks valuable information. 
We present privacy-preserving examples and conduct two in-depth user studies to evaluate the effectiveness and usability of \name{}.
Our results show that \name{} effectively protects the privacy of original visualizations. 


\begin{acks}
This project is supported by the Ministry of Education, Singapore, under its Academic Research Fund Tier 1 (NTU Tier 1, RG104/25). Any opinions, findings and conclusions or recommendations expressed in this material are those of the author(s) and do not reflect the views of the Ministry of Education, Singapore.
\end{acks}

\bibliographystyle{ACM-Reference-Format}
\bibliography{template}

@article{levi2008crowding,
  title={Crowding—An essential bottleneck for object recognition: A mini-review},
  author={Levi, Dennis M},
  journal={Vision research},
  volume={48},
  number={5},
  pages={635--654},
  year={2008},
  publisher={Elsevier}
}

@inproceedings{andow2017uiref,
  title={Uiref: analysis of sensitive user inputs in android applications},
  author={Andow, Benjamin and Acharya, Akhil and Li, Dengfeng and Enck, William and Singh, Kapil and Xie, Tao},
  booktitle={Proceedings of the 10th acm conference on security and privacy in wireless and mobile networks},
  pages={23--34},
  year={2017}
}

@inproceedings{huang2015supor,
  title={$\{$SUPOR$\}$: Precise and scalable sensitive user input detection for android apps},
  author={Huang, Jianjun and Li, Zhichun and Xiao, Xusheng and Wu, Zhenyu and Lu, Kangjie and Zhang, Xiangyu and Jiang, Guofei},
  booktitle={24th USENIX Security Symposium (USENIX Security 15)},
  pages={977--992},
  year={2015}
}

@book{few2013information,
  title={Information Dashboard Design: Displaying Data for At-a-glance Monitoring},
  author={Few, S.},
  isbn={9781938377006},
  url={https://books.google.com.sg/books?id=7k0EnAEACAAJ},
  year={2013},
  publisher={Analytics Press}
}

@incollection{blascheck2021characterizing,
  title={Characterizing glanceable visualizations: from perception to behavior change},
  author={Blascheck, Tanja and Bentley, Frank and Choe, Eun Kyoung and Horak, Tom and Isenberg, Petra},
  booktitle={Mobile Data Visualization},
  pages={151--176},
  year={2021},
  publisher={Chapman and Hall/CRC}
}

@article{Franz2000Grasping,title={Grasping Visual Illusions: No Evidence for a Dissociation Between Perception and Action},  author={V. Franz and K. Gegenfurtner and H. Bülthoff and M. Fahle},  journal={Psychological Science},  year={2000},  volume={11},  pages={20 - 25},  doi={10.1111/1467-9280.00209}  }

@book{ware2013information,
  title={Information Visualization: Perception for Design},
  author={Ware, C.},
  isbn={9780123814647},
  lccn={2012009489},
  series={Information Visualization: Perception for Design},
  url={https://books.google.com.sg/books?id=qFmS95vf6H8C},
  year={2013},
  publisher={Elsevier Science}
}

@article{gregory1997knowledge,
  title={Knowledge in perception and illusion},
  author={Gregory, Richard L},
  journal={Philosophical Transactions of the Royal Society of London. Series B: Biological Sciences},
  volume={352},
  number={1358},
  pages={1121--1127},
  year={1997},
  publisher={The Royal Society}
}

@article{andresyuk2024visual,
  title={Visual illusions as a means of aestheticisation of the object-spatial environment},
  author={Andresyuk, Borys},
  journal={Notes on Art Criticism},
  volume={1},
  number={24},
  pages={68--79},
  year={2024}
}

@incollection{kruglanski2018intuitive,
  title={Intuitive and deliberate judgments are based on common principles},
  author={Kruglanski, Arie W and Gigerenzer, Gerd},
  booktitle={The motivated mind},
  pages={104--128},
  year={2018},
  publisher={Routledge}
}

@inproceedings{abdrabou2022understanding,
  title={Understanding shoulder surfer behavior and attack patterns using virtual reality},
  author={Abdrabou, Yasmeen and Rivu, Sheikh Radiah and Ammar, Tarek and Liebers, Jonathan and Saad, Alia and Liebers, Carina and Gruenefeld, Uwe and Knierim, Pascal and Khamis, Mohamed and Makela, Ville and others},
  booktitle={Proceedings of the 2022 International Conference on Advanced Visual Interfaces},
  pages={1--9},
  year={2022}
}

@book{lee2021mobile,
  title={Mobile Data Visualization},
  author={Lee, Bongshin and Dachselt, Raimund and Isenberg, Petra and Choe, Eun Kyoung},
  year={2021},
  publisher={CRC Press}
}

@article{Kerren2008Information,title={Information Visualization},author={A. Kerren and John T. Stasko and Jean-Daniel Fekete},year={2008},pages={582},doi={10.1007/978-0-387-35973-1_639}}

@String{tog = "ACM TOG"}

@String{Springer = "Springer-Verlag" }

@String{Computing = "Computing" }

@String{Computer = "{IEEE} Computer" }

@article{oksanen2015methods,
  title={Methods for deriving and calibrating privacy-preserving heat maps from mobile sports tracking application data},
  author={Oksanen, Juha and Bergman, Cecilia and Sainio, Jani and Westerholm, Jan},
  journal={Journal of Transport Geography},
  volume={48},
  pages={135--144},
  year={2015},
  publisher={Elsevier}
}

@article{Chou2018PrivacyPV,
  title={Privacy Preserving Visualization: A Study on Event Sequence Data},
  author={Chou, Jia-Kai and Wang, Yang and Ma, Kwan Liu},
  journal={Computer Graphics Forum},
  year={2018},
  volume={38},
  url={https://api.semanticscholar.org/CorpusID:56947590}
}

@article{Chou2018AnES,
  title={An Empirical Study on Perceptually Masking Privacy in Graph Visualizations},
  author={Jia-Kai Chou and Chris Bryan and Jing Li and Kwan-Liu Ma},
  journal={2018 IEEE Symposium on Visualization for Cyber Security (VizSec)},
  year={2018},
  pages={1-8},
  url={https://api.semanticscholar.org/CorpusID:148573298}
}

@article{Panavas2023InvestigatingTV,
  title={Investigating the Visual Utility of Differentially Private Scatterplots.},
  author={Liudas Panavas and Tarik Crnovrsanin and J. L. Adams and Jonathan Ullman and Ali Sargavad and Melanie Tory and Cody Dunne},
  journal={IEEE Transactions on Visualization and Computer Graphics},
  year={2023},
  volume={PP},
  url={https://api.semanticscholar.org/CorpusID:259257746}
}

@article{Chen2020FederatedVA,
  title={Federated Visualization: A Privacy-Preserving Strategy for Aggregated Visual Query},
  author={Wei Chen and Yating Wei and Zhiyong Wang and Shuyue Zhou and Bingru Lin and Zhiguang Zhou},
  journal={IEEE Transactions on Visualization and Computer Graphics},
  year={2020},
  volume={29},
  pages={2901-2913},
  url={https://api.semanticscholar.org/CorpusID:246680117}
}

@inproceedings{dasgupta2013measuring,
  title={Measuring Privacy and Utility in Privacy-Preserving Visualization},
  author={Dasgupta, Aritra and Chen, Min and Kosara, Robert},
  booktitle={Computer Graphics Forum},
  volume={32},
  pages={35--47},
  year={2013},
  organization={Wiley Online Library}
}

@article{chou2016obfuscated,
  title={Obfuscated volume rendering},
  author={Chou, Jia-Kai and Yang, Chuan-Kai},
  journal={The Visual Computer},
  volume={32},
  number={12},
  pages={1593--1604},
  year={2016},
  publisher={Springer}
}

@inproceedings{dasgupta2019guess,
  title={Guess me if you can: A visual uncertainty model for transparent evaluation of disclosure risks in privacy-preserving data visualization},
  author={Dasgupta, Aritra and Kosara, Robert and Chen, Min},
  booktitle={Proceedings of the 2019 IEEE Symposium on Visualization for Cyber Security (VizSec)},
  pages={1--10},
  year={2019},
  organization={IEEE}
}

@inproceedings{chen2019keep,
  title={Keep others from peeking at your mobile device screen!},
  author={Chen, Chun-Yu and Lin, Bo-Yao and Wang, Junding and Shin, Kang G},
  booktitle={Proceedings of Annual International Conference on Mobile Computing and Networking},
  pages={1--16},
  year={2019}
}

@article{papadopoulos2017illusionpin,
  title={Illusionpin: Shoulder-surfing resistant authentication using hybrid images},
  author={Papadopoulos, Athanasios and Nguyen, Toan and Durmus, Emre and Memon, Nasir},
  journal={IEEE Transactions on Information Forensics and Security},
  volume={12},
  number={12},
  pages={2875--2889},
  year={2017},
  publisher={IEEE}
}

@inproceedings{tang2023eye,
  title={$\{$Eye-Shield$\}$:$\{$Real-Time$\}$ Protection of Mobile Device Screen Information from Shoulder Surfing},
  author={Tang, Brian Jay and Shin, Kang G},
  booktitle={32nd USENIX Security Symposium (USENIX Security 23)},
  pages={5449--5466},
  year={2023}
}

@inproceedings{von2016you,
  title={You Can't Watch This! Privacy-Respectful Photo Browsing on Smartphones},
  author={von Zezschwitz, Emanuel and Ebbinghaus, Sigrid and Hussmann, Heinrich and De Luca, Alexander},
  booktitle={Proceedings of the 2016 CHI Conference on Human Factors in Computing Systems},
  pages={4320--4324},
  year={2016}
}

@article{oliva2006hybrid,
  title={Hybrid images},
  author={Oliva, Aude and Torralba, Antonio and Schyns, Philippe G},
  journal={ACM Transactions on Graphics (TOG)},
  volume={25},
  number={3},
  pages={527--532},
  year={2006},
  publisher={ACM New York, NY, USA}
}

@book{munzner2014visualization,
  title={Visualization Analysis and Design},
  author={Munzner, Tamara},
  year={2014},
  publisher={CRC press}
}

@article{isenberg2013hybrid,
  title={Hybrid-image visualization for large viewing environments},
  author={Isenberg, Petra and Dragicevic, Pierre and Willett, Wesley and Bezerianos, Anastasia and Fekete, Jean-Daniel},
  journal={IEEE Transactions on Visualization \& Computer Graphics},
  volume={19},
  number={12},
  pages={2346--2355},
  year={2013},
  publisher={IEEE Computer Society}
}

@inproceedings{battle2018beagle,
  title={Beagle: Automated Extraction and Interpretation of Visualizations From the Web},
  author={Battle, Leilani and Duan, Peitong and Miranda, Zachery and Mukusheva, Dana and Chang, Remco and Stonebraker, Michael},
  booktitle={Proceedings of the 2018 CHI Conference on Human Factors in Computing Systems},
  pages={1--8},
  year={2018}
}

@inproceedings{ali2014protecting,
  title={Protecting mobile users from visual privacy attacks},
  author={Ali, Mohammed Eunus and Anwar, Anika and Ahmed, Ishrat and Hashem, Tanzima and Kulik, Lars and Tanin, Egemen},
  booktitle={Proceedings of the 2014 ACM International Joint Conference on Pervasive and Ubiquitous Computing: Adjunct Publication},
  pages={1--4},
  year={2014}
}

@inproceedings{eiband2017understanding,
  title={Understanding shoulder surfing in the wild: Stories from users and observers},
  author={Eiband, Malin and Khamis, Mohamed and Von Zezschwitz, Emanuel and Hussmann, Heinrich and Alt, Florian},
  booktitle={Proceedings of the 2017 CHI Conference on Human Factors in Computing Systems},
  pages={4254--4265},
  year={2017}
}

@inproceedings{dasgupta2012conceptualizing,
  title={Conceptualizing visual uncertainty in parallel coordinates},
  author={Dasgupta, Aritra and Chen, Min and Kosara, Robert},
  booktitle={Computer Graphics Forum},
  volume={31},
  pages={1015--1024},
  year={2012},
  organization={Wiley Online Library}
}

@article{wang2018graphprotector,
  title={Graphprotector: A visual interface for employing and assessing multiple privacy preserving graph algorithms},
  author={Wang, Xumeng and Chen, Wei and Chou, Jia-Kai and Bryan, Chris and Guan, Huihua and Chen, Wenlong and Pan, Rusheng and Ma, Kwan-Liu},
  journal={IEEE Transactions on Visualization and Computer Graphics},
  volume={25},
  number={1},
  pages={193--203},
  year={2018},
  publisher={IEEE}
}

@article{yoshimura2017smartphone,
  title={Smartphone viewing distance and sleep: an experimental study utilizing motion capture technology},
  author={Yoshimura, Michitaka and Kitazawa, Momoko and Maeda, Yasuhiro and Mimura, Masaru and Tsubota, Kazuo and Kishimoto, Taishiro},
  journal={Nature and Science of Sleep},
  volume={9},
  pages={59},
  year={2017},
  publisher={Dove Press}
}

@misc{
SeatWidth2021,
url={https://www.kosokubus.com/en/special/expressbus.html}, 
key={Kosokubus},
year={2023}, 
month={Oct}
}

@article{senay1994knowledge,
  title={A knowledge-based system for visualization design},
  author={Senay, Hikmet and Ignatius, Eve},
  journal={IEEE Computer Graphics and Applications},
  volume={14},
  number={6},
  pages={36--47},
  year={1994},
  publisher={IEEE}
}

@inproceedings{wiedenbeck2006design,
  title={Design and evaluation of a shoulder-surfing resistant graphical password scheme},
  author={Wiedenbeck, Susan and Waters, Jim and Sobrado, Leonardo and Birget, Jean-Camille},
  booktitle={Proceedings of the Working Conference on Advanced Visual Interfaces},
  pages={177--184},
  year={2006}
}

@article{Sarathy2023DontLA,
  title={Don't Look at the Data! How Differential Privacy Reconfigures the Practices of Data Science},
  author={Jayshree Sarathy and Sophia Song and Audrey Haque and Tania Schlatter and Salil P. Vadhan},
  journal={ArXiv},
  year={2023},
  volume={abs/2302.11775}
}

@inproceedings{dekel2017effectiveness,
  title={On the effectiveness of visible watermarks},
  author={Dekel, Tali and Rubinstein, Michael and Liu, Ce and Freeman, William T},
  booktitle={Proceedings of the IEEE Conference on Computer Vision and Pattern Recognition},
  pages={2146--2154},
  year={2017}
}

@inproceedings{mohanty2000dct,
  title={A DCT domain visible watermarking technique for images},
  author={Mohanty, Saraju P and Ramakrishnan, Kalpathi R and Kankanhalli, Mohan S},
  booktitle={Proceedings of the 2000 IEEE International Conference on Multimedia and Expo. ICME2000. Latest Advances in the Fast Changing World of Multimedia (Cat. No. 00TH8532)},
  volume={2},
  pages={1029--1032},
  year={2000},
  organization={IEEE}
}

@article{begum2020digital,
  title={Digital image watermarking techniques: a review},
  author={Begum, Mahbuba and Uddin, Mohammad Shorif},
  journal={Information},
  volume={11},
  number={2},
  pages={110},
  year={2020},
  publisher={MDPI}
}

@inproceedings{braudaway1997protecting,
  title={Protecting publicly-available images with an invisible image watermark},
  author={Braudaway, Gordon W},
  booktitle={Proceedings of the International Conference on Image Processing},
  volume={1},
  pages={524--527},
  year={1997},
  organization={IEEE}
}

@inproceedings{meng1998embedding,
  title={Embedding visible video watermarks in the compressed domain},
  author={Meng, Jianhao and Chang, Shih-Fu},
  booktitle={Proceedings of the 1998 International Conference on Image Processing. ICIP98 (Cat. No. 98CB36269)},
  volume={1},
  pages={474--477},
  year={1998},
  organization={IEEE}
}

@article{hu2001wavelet,
  title={Wavelet domain adaptive visible watermarking},
  author={Hu, Yongjian and Kwong, Sam},
  journal={Electronics Letters},
  volume={37},
  number={20},
  pages={1},
  year={2001},
  publisher={The Institution of Engineering \& Technology}
}

@article{huang2006contrast,
  title={A contrast-sensitive visible watermarking scheme},
  author={Huang, Biao-Bing and Tang, Shao-Xian},
  journal={IEEE MultiMedia},
  volume={13},
  number={2},
  pages={60--66},
  year={2006},
  publisher={IEEE}
}

@article{connor2004visual,
  title={Visual attention: bottom-up versus top-down},
  author={Connor, Charles E and Egeth, Howard E and Yantis, Steven},
  journal={Current Biology},
  volume={14},
  number={19},
  pages={R850--R852},
  year={2004},
  publisher={Elsevier}
}

@article{nothdurft2000salience,
  title={Salience from feature contrast: additivity across dimensions},
  author={Nothdurft, Hans-Christoph},
  journal={Vision Research},
  volume={40},
  number={10-12},
  pages={1183--1201},
  year={2000},
  publisher={Elsevier}
}

@article{terzic2017texture,
  title={Texture features for object salience},
  author={Terzi{\'c}, Kasim and Krishna, Sai and du Buf, JM Hans},
  journal={Image and Vision Computing},
  volume={67},
  pages={43--51},
  year={2017},
  publisher={Elsevier}
}

@article{white2017colour,
  title={Colour and luminance contrasts predict the human detection of natural stimuli in complex visual environments},
  author={White, Thomas E and Rojas, Bibiana and Mappes, Johanna and Rautiala, Petri and Kemp, Darrell J},
  journal={Biology Letters},
  volume={13},
  number={9},
  pages={20170375},
  year={2017},
  publisher={The Royal Society}
}

@article{gilbert2007brain,
  title={Brain states: top-down influences in sensory processing},
  author={Gilbert, Charles D and Sigman, Mariano},
  journal={Neuron},
  volume={54},
  number={5},
  pages={677--696},
  year={2007},
  publisher={Elsevier}
}

@article{matzen2017data,
  title={Data visualization saliency model: A tool for evaluating abstract data visualizations},
  author={Matzen, Laura E and Haass, Michael J and Divis, Kristin M and Wang, Zhiyuan and Wilson, Andrew T},
  journal={IEEE Transactions on Visualization and Computer Graphics},
  volume={24},
  number={1},
  pages={563--573},
  year={2017},
  publisher={IEEE}
}

@article{rosli2015gestalt,
  title={Gestalt principles in multimodal data representation},
  author={Rosli, Muhammad Hafiz Wan and Cabrera, Andres},
  journal={IEEE Computer Graphics and Applications},
  volume={35},
  number={2},
  pages={80--87},
  year={2015},
  publisher={IEEE}
}

@article{huber1982adding,
  title={Adding asymmetrically dominated alternatives: Violations of regularity and the similarity hypothesis},
  author={Huber, Joel and Payne, John W and Puto, Christopher},
  journal={Journal of consumer research},
  volume={9},
  number={1},
  pages={90--98},
  year={1982},
  publisher={The University of Chicago Press}
}

@article{yuen1990comparative,
  title={Comparative study of Hough transform methods for circle finding},
  author={Yuen, HK and Princen, John and Illingworth, John and Kittler, Josef},
  journal={Image and vision computing},
  volume={8},
  number={1},
  pages={71--77},
  year={1990},
  publisher={Elsevier}
}

@article{leavers1993hough,
  title={Which hough transform?},
  author={Leavers, VF},
  journal={CVGIP: Image Understanding},
  volume={58},
  number={2},
  pages={250--264},
  year={1993},
  publisher={Elsevier}
}

@article{suzuki1985topological,
  title={Topological structural analysis of digitized binary images by border following},
  author={Suzuki, Satoshi and others},
  journal={Computer Vision, Graphics, and Image Processing},
  volume={30},
  number={1},
  pages={32--46},
  year={1985},
  publisher={Elsevier}
}

@inproceedings{heer2010crowdsourcing,
  title={Crowdsourcing graphical perception: using mechanical turk to assess visualization design},
  author={Heer, Jeffrey and Bostock, Michael},
  booktitle={Proceedings of the SIGCHI conference on human factors in computing systems},
  pages={203--212},
  year={2010}
}

@article{zhang2023don,
  title={Don't Peek at My Chart: Privacy-preserving Visualization for Mobile Devices},
  author={Zhang, Songheng and Ma, Dong and Wang, Yong},
  journal={arXiv preprint arXiv:2303.13307},
  year={2023}
}

@article{berns1993mathematical,
  title={The mathematical development of CIE TC 1-29 proposed color difference equation: CIELCH},
  author={Berns, Roy S},
  journal={Proceedings of AIC colour},
  volume={93},
  pages={189--192},
  year={1993}
}

@incollection{HUSSEIN201197,
title = {9 - Preprocessing of Measurements},
editor = {Esam M.A. Hussein},
booktitle = {Computed Radiation Imaging},
publisher = {Elsevier},
address = {London},
pages = {97-123},
year = {2011},
isbn = {978-0-12-387777-2},
doi = {https://doi.org/10.1016/B978-0-12-387777-2.00009-4},
url = {https://www.sciencedirect.com/science/article/pii/B9780123877772000094},
author = {Esam M.A. Hussein}
}

@inproceedings{gu2013self,
  title={Self-adaptive scale transform for IQA metric},
  author={Gu, Ke and Zhai, Guangtao and Yang, Xiaokang and Zhang, Wenjun},
  booktitle={Proceedings of the 2013 IEEE International Symposium on Circuits and Systems (ISCAS)},
  pages={2365--2368},
  year={2013},
  organization={IEEE}
}

@article{gu2015quality,
  title={Quality assessment considering viewing distance and image resolution},
  author={Gu, Ke and Liu, Min and Zhai, Guangtao and Yang, Xiaokang and Zhang, Wenjun},
  journal={IEEE Transactions on Broadcasting},
  volume={61},
  number={3},
  pages={520--531},
  year={2015},
  publisher={IEEE}
}

@inproceedings{wang2003multiscale,
  title={Multiscale structural similarity for image quality assessment},
  author={Wang, Zhou and Simoncelli, Eero P and Bovik, Alan C},
  booktitle={Proceedings of the Thirty-Seventh Asilomar Conference on Signals, Systems \& Computers, 2003},
  volume={2},
  pages={1398--1402},
  year={2003},
  organization={Ieee}
}

@article{zhang2014vsi,
  title={VSI: A visual saliency-induced index for perceptual image quality assessment},
  author={Zhang, Lin and Shen, Ying and Li, Hongyu},
  journal={IEEE Transactions on Image Processing},
  volume={23},
  number={10},
  pages={4270--4281},
  year={2014},
  publisher={IEEE}
}

@online{plotly, author = {Plotly Technologies Inc.}, title = {Collaborative data science}, publisher = {Plotly Technologies Inc.}, address = {Montreal, QC}, year = {2015}, url = {https://plot.ly} }

@inproceedings{lee2018data,
  title={Data visualization on mobile devices},
  author={Lee, Bongshin and Brehmer, Matthew and Isenberg, Petra and Choe, Eun Kyoung and Langner, Ricardo and Dachselt, Raimund},
  booktitle={Proceedings of Extended Abstracts of the 2018 CHI Conference on Human Factors in Computing Systems},
  pages={1--8},
  year={2018}
}

@article{panetta2008human,
  title={Human visual system-based image enhancement and logarithmic contrast measure},
  author={Panetta, Karen A and Wharton, Eric J and Agaian, Sos S},
  journal={IEEE Transactions on Systems, Man, and Cybernetics, Part B (Cybernetics)},
  volume={38},
  number={1},
  pages={174--188},
  year={2008},
  publisher={IEEE}
}

@book{devalois1990spatial,
  title={Spatial vision},
  author={DeValois, Russell L and Burke, Phyllis and De Valois, Karen K and DeValois, Karen K},
  year={1990},
  publisher={Oxford University Press on Demand}
}

@article{rovamo1992contrast,
  title={Contrast sensitivity as a function of spatial frequency, viewing distance and eccentricity with and without spatial noise},
  author={Rovamo, Jyrki and Franssila, Rauli and N{\"a}s{\"a}nen, Risto},
  journal={Vision Research},
  volume={32},
  number={4},
  pages={631--637},
  year={1992},
  publisher={Elsevier}
}

@article{brehmer2019comparative,
  title={A comparative evaluation of animation and small multiples for trend visualization on mobile phones},
  author={Brehmer, Matthew and Lee, Bongshin and Isenberg, Petra and Choe, Eun Kyoung},
  journal={IEEE Transactions on Visualization and Computer Graphics},
  volume={26},
  number={1},
  pages={364--374},
  year={2019},
  publisher={IEEE}
}

@article{zeng2023semi,
  title={Semi-Automatic Layout Adaptation for Responsive Multiple-View Visualization Design},
  author={Zeng, Wei and Chen, Xi and Hou, Yihan and Shao, Lingdan and Chu, Zhe and Chang, Remco},
  journal={IEEE Transactions on Visualization and Computer Graphics},
  year={2023},
  publisher={IEEE}
}

@article{todorovic2008gestalt,
  title={Gestalt principles},
  author={Todorovic, Dejan},
  journal={Scholarpedia},
  volume={3},
  number={12},
  pages={5345},
  year={2008}
}

@article{borland2007rainbow,
  title={Rainbow color map (still) considered harmful},
  author={Borland, David and Ii, Russell M Taylor},
  journal={IEEE Computer Graphics and Applications},
  volume={27},
  number={2},
  pages={14--17},
  year={2007},
  publisher={IEEE}
}

@article{dastani2002role,
  title={The role of visual perception in data visualization},
  author={Dastani, Mehdi},
  journal={Journal of Visual Languages \& Computing},
  volume={13},
  number={6},
  pages={601--622},
  year={2002},
  publisher={Elsevier}
}

@inproceedings{bartram2017affective,
  title={Affective color in visualization},
  author={Bartram, Lyn and Patra, Abhisekh and Stone, Maureen},
  booktitle={Proceedings of the 2017 CHI conference on human factors in computing systems},
  pages={1364--1374},
  year={2017}
}

@inproceedings{nesbitt2002applying,
  title={Applying gestalt principles to animated visualizations of network data},
  author={Nesbitt, Keith V and Friedrich, Carsten},
  booktitle={Proceedings the Sixth International Conference on Information Visualisation},
  pages={737--743},
  year={2002},
  organization={IEEE}
}

@article{tran2021approaching,
  title={Approaching human vision perception to designing visual graph in data visualization},
  author={Tran, Phuoc Vinh and Le, Truong Xuan},
  journal={Concurrency and Computation: Practice and Experience},
  volume={33},
  number={2},
  pages={e5722},
  year={2021},
  publisher={Wiley Online Library}
}

@book{jammalamadaka2001topics,
  title={Topics in circular statistics},
  author={Jammalamadaka, S Rao and SenGupta, Ashis},
  volume={5},
  year={2001},
  publisher={world scientific}
}

@article{levin1993visual,
  title={Visual angle as a determinant of perceived interobject distance},
  author={Levin, Charles A and Haber, Ralph Norman},
  journal={Perception \& Psychophysics},
  volume={54},
  number={2},
  pages={250--259},
  year={1993},
  publisher={Springer}
}

@book{cliff2014ordinal,
  title={Ordinal methods for behavioral data analysis},
  author={Cliff, Norman},
  year={2014},
  publisher={Psychology Press}
}

@book{iliinsky2011designing,
  title={Designing data visualizations: Representing informational Relationships},
  author={Iliinsky, Noah and Steele, Julie},
  year={2011},
  publisher={" O'Reilly Media, Inc."}
}

@article{Herrmann2010WhenSM,
  title={When size matters: attention affects performance by contrast or response gain},
  author={Katrin Herrmann and Leila Montaser-Kouhsari and Marisa Carrasco and David J. Heeger},
  journal={Nature neuroscience},
  year={2010},
  volume={13},
  pages={1554 - 1559},
  url={https://api.semanticscholar.org/CorpusID:10197681}
}

@article{Stuart2003TheOI,
  title={The overlay interference task and object-selective visual attention},
  author={Geoffrey William Stuart and Ken I. McAnally and James W. Meehan},
  journal={Vision Research},
  year={2003},
  volume={43},
  pages={1443-1453},
  url={https://api.semanticscholar.org/CorpusID:14725285}
}

@article{duda1972use,
  title={Use of the Hough transformation to detect lines and curves in pictures},
  author={Duda, Richard O and Hart, Peter E},
  journal={Communications of the ACM},
  volume={15},
  number={1},
  pages={11--15},
  year={1972},
  publisher={ACM New York, NY, USA}
}

@inproceedings{Pedersen2009CircularHT,
  title={Circular Hough Transform},
  author={Simon Just Kjeldgaard Pedersen},
  booktitle={Encyclopedia of Biometrics},
  year={2009},
  url={https://api.semanticscholar.org/CorpusID:17799110}
}

@article{cleveland1984graphical,
  title={Graphical perception: Theory, experimentation, and application to the development of graphical methods},
  author={Cleveland, William S and McGill, Robert},
  journal={Journal of the American statistical association},
  volume={79},
  number={387},
  pages={531--554},
  year={1984},
  publisher={Taylor \& Francis}
}

@article{Wang2018AVF,
  title={A Vector Field Design Approach to Animated Transitions},
  author={Yong Wang and Daniel Archambault and Carlos Eduardo Scheidegger and Huamin Qu},
  journal={IEEE Transactions on Visualization and Computer Graphics},
  year={2018},
  volume={24},
  pages={2487-2500},
  url={https://api.semanticscholar.org/CorpusID:206806433}
}

@article{Bce2022PrivacyScoutAV,
  title={PrivacyScout: Assessing Vulnerability to Shoulder Surfing on Mobile Devices},
  author={Mihai B{\^a}ce and Alia Saad and M. Khamis and Stefan Schneegass and Andreas Bulling},
  journal={Proc. Priv. Enhancing Technol.},
  year={2022},
  volume={2022},
  pages={650-669},
  url={https://api.semanticscholar.org/CorpusID:249943346}
}

@incollection{hart1988development,
  title={Development of NASA-TLX (Task Load Index): Results of empirical and theoretical research},
  author={Hart, Sandra G and Staveland, Lowell E},
  booktitle={Advances in psychology},
  volume={52},
  pages={139--183},
  year={1988},
  publisher={Elsevier}
}

@article{nygren1991psychometric,
  title={Psychometric properties of subjective workload measurement techniques: Implications for their use in the assessment of perceived mental workload},
  author={Nygren, Thomas E},
  journal={Human factors},
  volume={33},
  number={1},
  pages={17--33},
  year={1991},
  publisher={SAGE Publications Sage CA: Los Angeles, CA}
}

@article{lewis1995ibm,
  title={IBM computer usability satisfaction questionnaires: psychometric evaluation and instructions for use},
  author={Lewis, James R},
  journal={International Journal of Human-Computer Interaction},
  volume={7},
  number={1},
  pages={57--78},
  year={1995},
  publisher={Taylor \& Francis}
}

\newpage
\onecolumn
\appendix
\section{More Privacy-preserving Vis. Examples}
\label{appendix: examples}

\begin{figure*}[!htb]
    \centering
    \includegraphics[width=\linewidth]{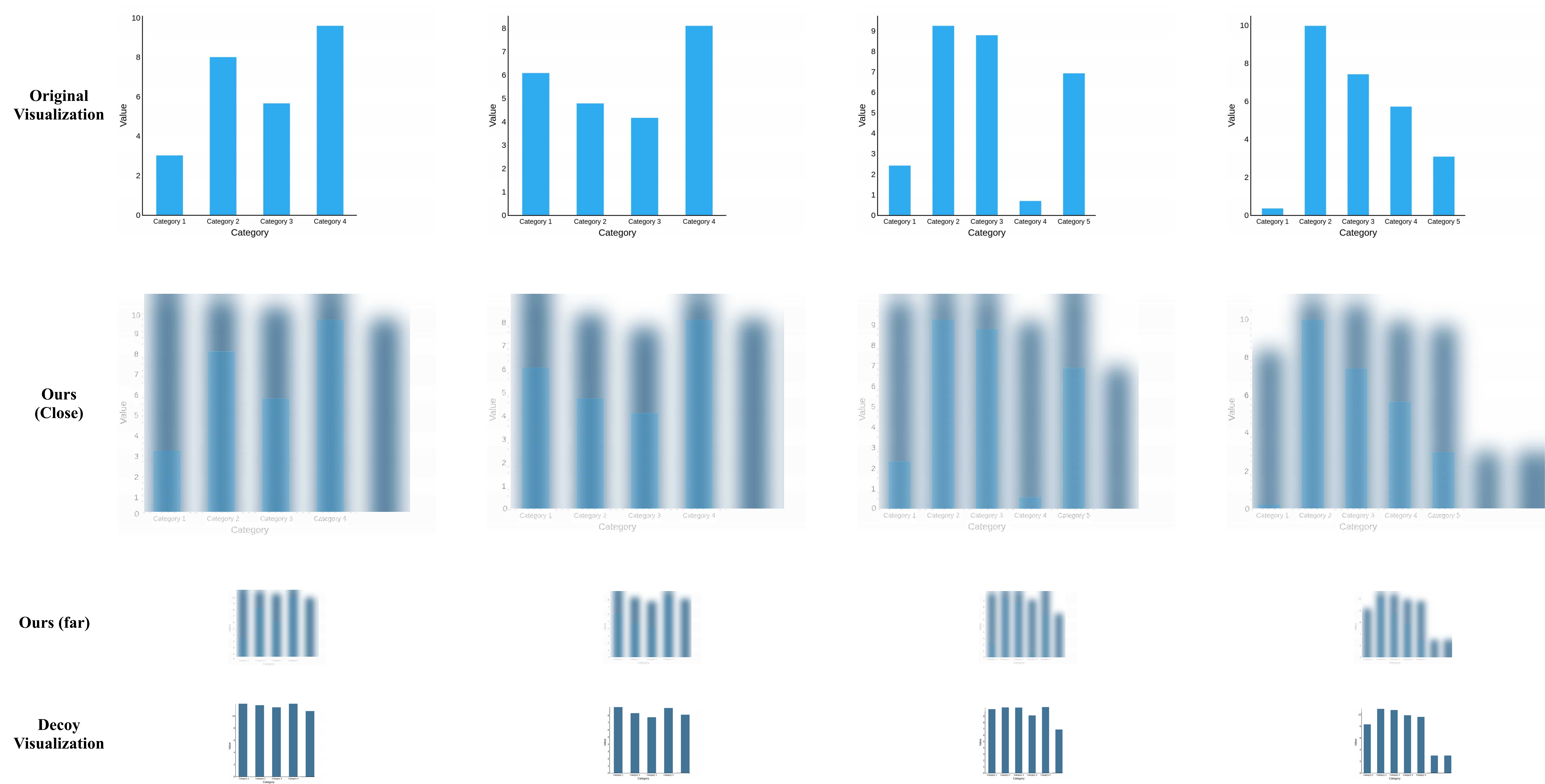}
    \caption{An illustration of privacy-preserving bar charts generated by \name{}. The figure showcases four distinct bar charts (four columns), displaying the simulated perception of original visualizations at a close distance (top row), privacy-preserving visualizations at a close distance (second row), privacy-preserving visualizations at a far distance (third row), and decoy visualizations at a far distance (bottom row).  \textbf{Note}: the effect of privacy-preserving visualizations (second row) depends on their displayed sizes, and \textbf{readers are recommended to expand them to validate their effectiveness if they are displayed in a small size}.}
\label{fig: appendix_bar}
\end{figure*}

\begin{figure*}[!hbt]
    \centering
    \includegraphics[width=\linewidth]{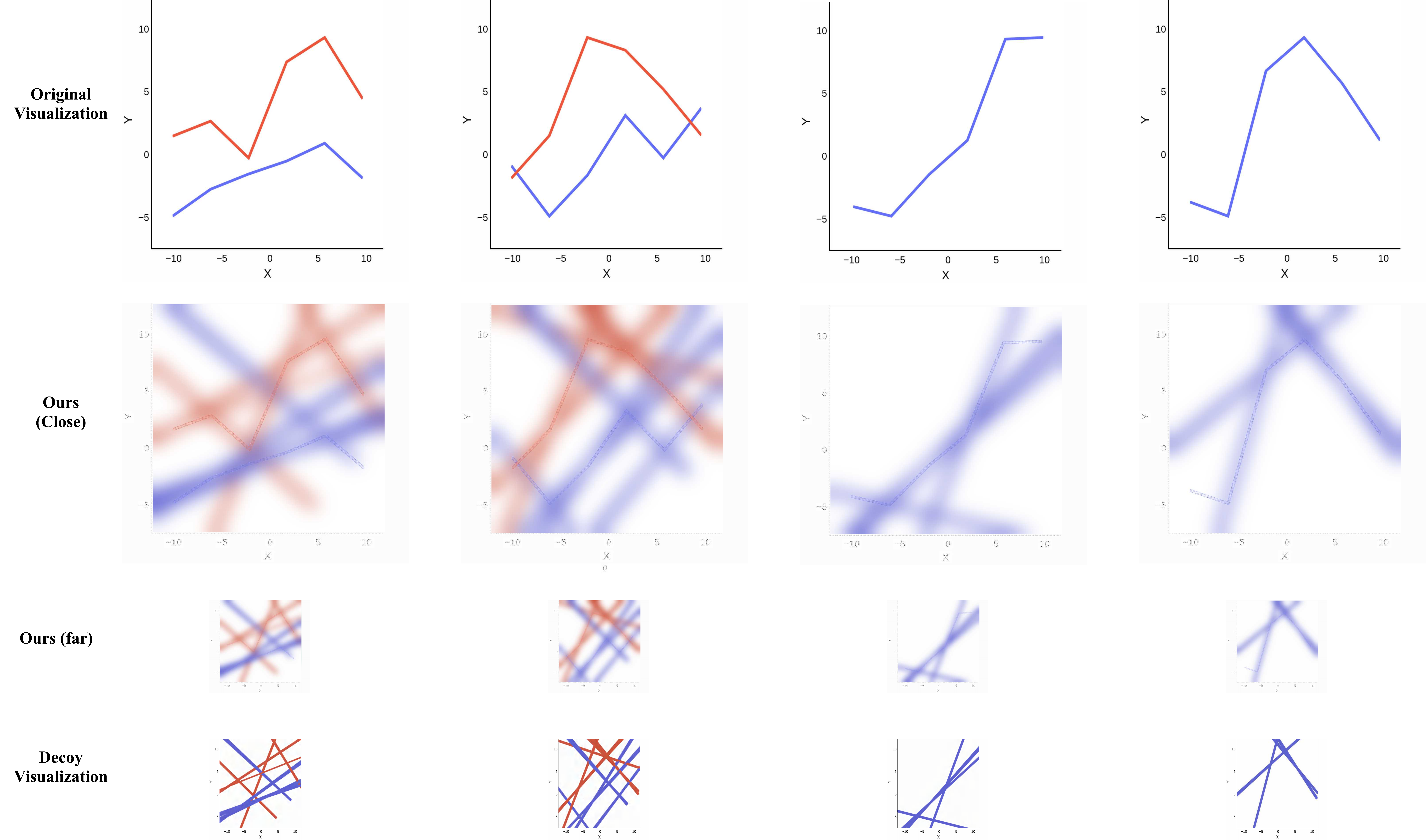}
    \caption{An illustration of privacy-preserving line charts generated by \name{}. The figure demonstrates four distinct line chart (four columns), displaying the simulated perception of original visualizations at a close distance (top row), privacy-preserving visualizations at a close distance (second row), privacy-preserving visualizations at a far distance (third row), and decoy visualizations at a far distance (bottom row).  \textbf{Note}: the effect of privacy-preserving visualizations (second row) depends on their displayed sizes, and \textbf{readers are recommended to expand them to validate their effectiveness if they are displayed in a small size}.}
\label{fig: appendix_line}

\end{figure*}

\begin{figure*}[hbt!]
    \centering
    \includegraphics[width=\linewidth]{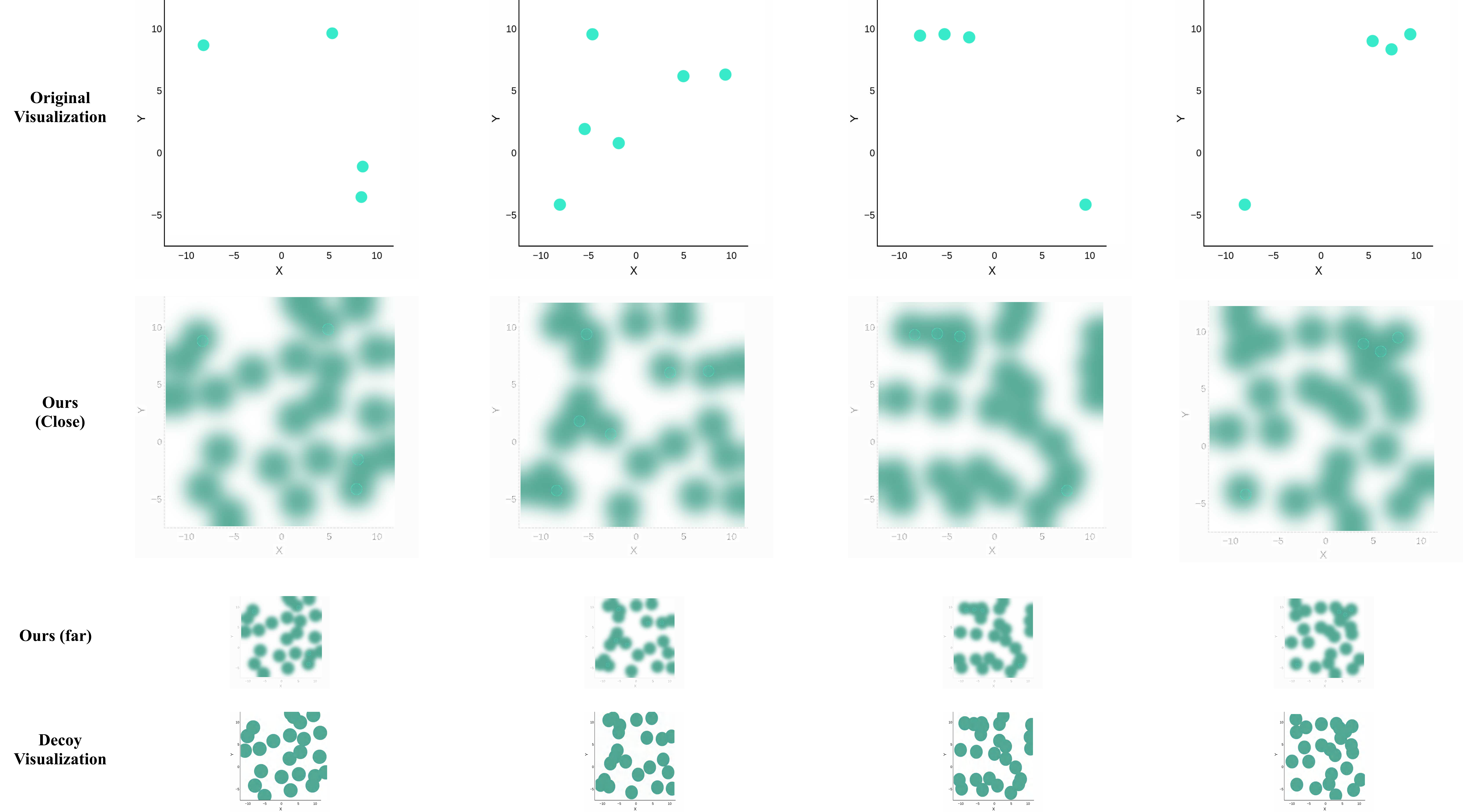}
    \caption{An illustration of privacy-preserving scatter plots generated by \name{}. The figure showcases four distinct scatter plots (four columns), displaying the simulated perception of original visualizations at a close distance (top row), privacy-preserving visualizations at a close distance (second row), privacy-preserving visualizations at a far distance (third row), and decoy visualizations at a far distance (bottom row).  \textbf{Note}: the effect of privacy-preserving visualizations (second row) depends on their displayed sizes, and \textbf{readers are recommended to expand them to validate their effectiveness if they are displayed in a small size}.}
\label{fig: appendix_scatter}

\end{figure*}

\begin{figure*}[hbt!]
    \centering
    \includegraphics[width=\linewidth]{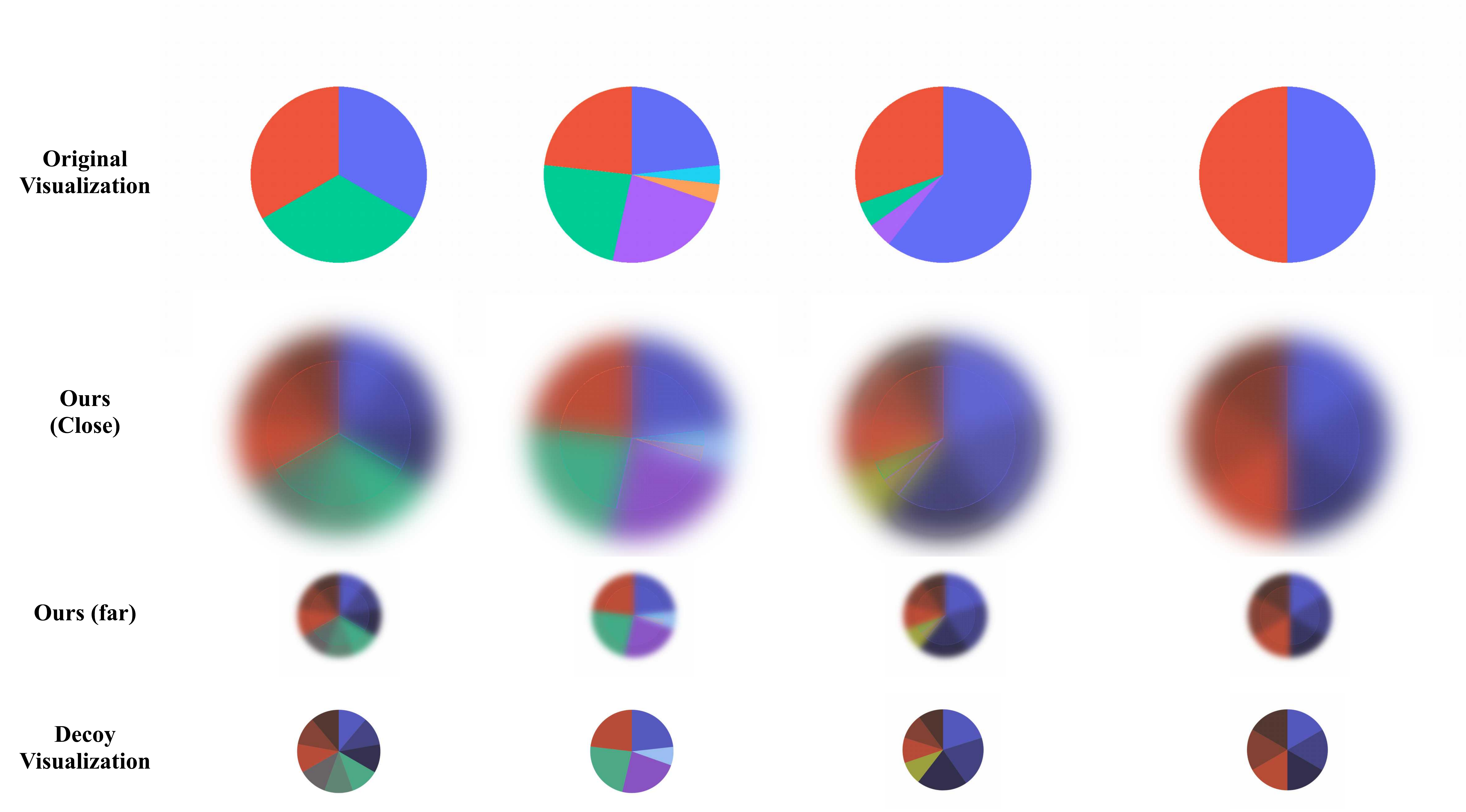}
    \caption{An illustration of privacy-preserving pie charts generated using our proposed \name{}. The figure showcases four different pie charts (four columns), displaying the simulated perception of original visualizations at a close distance (top row), privacy-preserving visualizations at a close distance (second row), privacy-preserving visualizations at a far distance (third row), and decoy visualizations at a far distance (bottom row). \textbf{Note}: the effect of privacy-preserving visualizations (second row) depends on their displayed sizes, and \textbf{readers are recommended to expand them to validate their effectiveness if they are displayed in a small size}.}
\label{fig: appendix_pie}

\end{figure*}

\begin{figure*}[ht!]
    \centering
    \includegraphics[width=\linewidth]{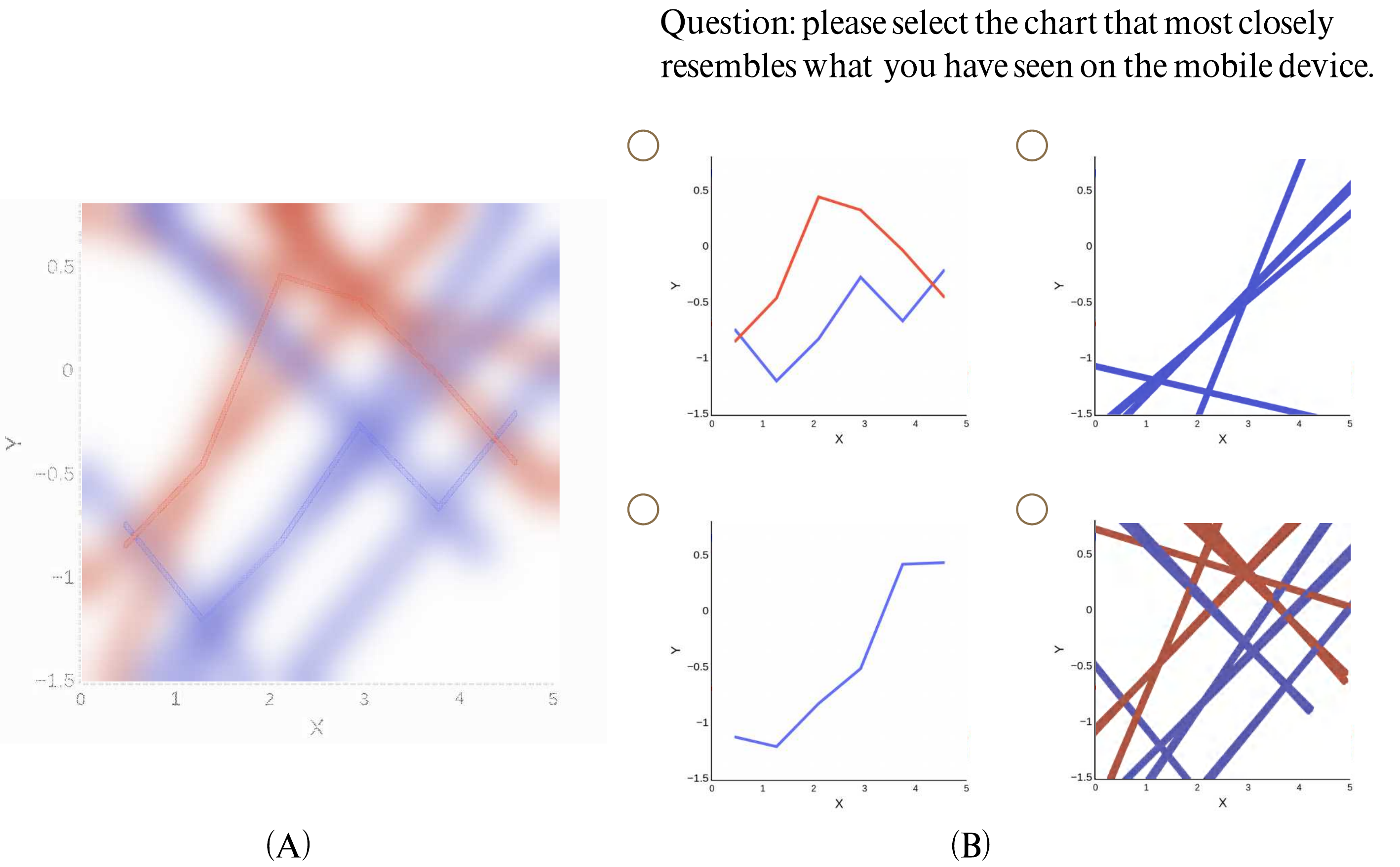}
    \caption{An illustration of a question example utilized in our user study for Real Users (RUs) or Shoulder Surfers (SSs). (A) An example of privacy-preserving visualization generated by \name{}. (B) An instance of a question related to a privacy-preserving visualization.}
\label{fig: appendix_example}

\end{figure*}

In the appendix, we provide more examples of privacy-preserving visualizations, demonstrating the effectiveness of our method across four common chart types. These examples underscore the versatility and adaptability of our approach in addressing privacy concerns for different visualizations. Figure~\ref{fig: appendix_bar} illustrates privacy-preserving bar chart examples, highlighting the capability of our method to effectively preserve privacy while maintaining the visual integrity of the original chart. Figure~\ref{fig: appendix_line} presents examples of privacy-preserving line charts, emphasizing the strength of our method in preserving privacy for visualizations that display continuous data over continuous data. In Figure~\ref{fig: appendix_scatter}, we depict examples of privacy-preserving scatter plots, showcasing the efficacy of our method in handling visualizations that represent relationships between variables. Lastly, Figure~\ref{fig: appendix_pie} provides examples of privacy-preserving pie charts, demonstrating the ability of our method to maintain privacy while preserving the visual clarity of the original chart. Figure~\ref{fig: appendix_example} depicts a question example in the user study.

\section{Optimization Model Details}\label{app:optimization}

This appendix provides the detailed formulas and parameter values for the perception-driven combinatorial optimization model described in Section~4.

\subsection{Distance-dependent Visualization Rescaling}
The downsampling factor $\gamma$ is correlated with the viewing distance $D$. It is calculated by first determining the human visual height ($H_v$) and width ($W_v$) based on human horizontal ($\theta_H = 40^\circ$) and vertical ($\theta_W = 50^\circ$) visual angles~\cite{gu2015quality}:

\begin{equation}
    \begin{aligned}
        H_{v} = 2\cdot\text{tan}(\frac{\theta_{H}}{2}) \cdot D,
    \end{aligned}
\end{equation}
\begin{equation}
    \begin{aligned}
        W_{v} = 2\cdot\text{tan}(\frac{\theta_{W}}{2}) \cdot D,
    \end{aligned}
\end{equation}

The downsampling factor $\gamma$ is then calculated using the visualization's actual height ($H_i$) and width ($W_i$):

\begin{equation}
    \begin{aligned}
       \gamma = \sqrt{\frac{H_{i}\cdot W_{i}}{H_{v} \cdot W_{v}}},
    \end{aligned}
\end{equation}

\subsection{Score Function}
The $\text{Score}$ function is designed to maximize the perceptual difference between close and far viewing distances. It is a weighted sum of two gap values.

$\text{Gap}_{1}$ models the visibility of the original visualization, defined as the difference in VSI similarity between close and far distances:
\begin{equation}
\begin{aligned}
\text{Gap}_{1} &= F_{V}(I^{o}_{c}, I^{p}_{c}) - F_{V}(I^{o}_{f}, I^{p}_{f}),
\end{aligned}
\end{equation}
where $F_{V}$ is the Visual Saliency-based Index (VSI)~\cite{zhang2014vsi}.

$\text{Gap}_{2}$ models the visibility of the decoy visualization, defined as the difference in MS-SSIM similarity between far and close distances:
\begin{equation}
\begin{aligned}
\text{Gap}_{2} &= F_{M}(I^{d}_{f}, I^{p}_{f}) - F_{M}(I^{d}_{c}, I^{p}_{c}),
\end{aligned}
\end{equation}
where $F_{M}$ is the Multi-Scale Structural Similarity (MS-SSIM) index~\cite{wang2003multiscale}.

The final score is the weighted sum of these two gaps. We empirically set $\alpha = 0.5$ and $\beta = 0.5$:
\begin{equation}
\begin{aligned}
\text{Score} &= \alpha \cdot \text{Gap}_{1} + \beta \cdot \text{Gap}_{2},
\end{aligned}
\end{equation}

\subsection{Exhaustive Search Parameter Ranges}
The \textit{exhaustive search} evaluates combinations of the following discretized values:
\begin{itemize}
\item Luminance: \{0, 0.1, 0.2, 0.3, ..., 100\},
\item Chroma: \{0, 0.1, 0.2, 0.3, ..., 100\},
\item Kernel Size:\{1, 3, 5, ..., $\text{min}(I_{w}, I_{h})$\},
\item Mask Area: \{1, 2, 3, ..., $\text{min}(V_{w}, V_{h})$\}.
\end{itemize}
Here, 0 and 100 are the min/max L and C channel values. $\text{min}(I_{w}, I_{h})$ is the minimum of the visualization's width and height, and $\text{min}(V_{w}, V_{h})$ is the minimum of the visual elements' width and height.

\end{document}